\documentclass[10pt]{article}

\usepackage{amsfonts,amssymb,amsmath,mathrsfs,latexsym,bbm,dsfont,amsthm}
\usepackage{color}
\usepackage{graphicx}

\usepackage[scriptsize,nooneline,hang]{caption}
\usepackage[hang,nooneline,scriptsize]{subfigure}

\usepackage{float}

\usepackage{authblk}

\allowdisplaybreaks 

\newcommand{\half}{\frac{1}{2}}
\newcommand{\dd}{{\rm d}}

\begin{document}

\title{Charged dilatonic black rings and black saturns\\ and their thermodynamics}

\author{Saskia Grunau}

\affil{Institut f\"ur Physik, Universit\"at Oldenburg, D--26111 Oldenburg, Germany}

\maketitle

\begin{abstract}
In this paper charged black rings and black saturns are constructed in Einstein-Maxwell-dilaton theory in five dimensions. The neutral black ring and black saturn solutions are embedded in six dimensions and boosted with respect to the time coordinate and the added sixth dimension. Then the charged solutions are obtained by a Kaluza-Klein reduction.
The influence of the charge is studied by analysing the physical properties and the phase diagram. The different dilatonic solutions are compared and their thermodynamic stability is considered.
\end{abstract}

\section{Introduction}

A promising development in the search for a quantum theory of gravity is string theory, which needs more than four dimensions for its internal consistency. This resulted in wide interest in higher-dimensional solutions including higher-dimensional black holes (see e.g.~\cite{Emparan:2008eg}). Myers and Perry found the higher-dimensional version of the Kerr black hole \cite{Myers:1986un} and also supposed the existence of black holes in higher dimensions with a nonspherical horizon topology. A solution with horizon topology $S^1 \times S^2$, a black ring, was discovered by Emparan and Reall \cite{Emparan:2001wn}. This showed that the uniqueness theorem does not hold in higher dimensions.
 
Since then a variety of black ring solutions has been found, e.g.~a black ring with two angular momenta \cite{Pomeransky:2006bd}, a charged black ring  \cite{Elvang:2003yy}, and a supersymmetric black ring \cite{Elvang:2004rt,Elvang:2004ds}. Also composite black objects are possible: a black saturn, a spherical black hole surrounded by a black ring \cite{Elvang:2007rd}, systems of multiple black rings \cite{Iguchi:2007is, Evslin:2007fv, Elvang:2007hs}, or black saturns with multiple rings.

In the static case, magnetic \cite{Yazadjiev:2008ty} and electric charge \cite{Chng:2008sr} have been added to the black saturn solution. The magnetized black saturn is balanced, but the  electrically charged black saturn exhibits either a conical or a naked singularity.\\

When gravity is coupled to a Maxwell field and a dilaton field, charged dilatonic black hole solutions emerge. In the case of Kaluza-Klein coupling the solutions can be obtained by adding a spatial dimension and performing a boost. A dimensional reduction then leads to the charged dilatonic black hole solutions (compare \cite{Maison:1979kx} -- \cite{Kunz:2006jd}).

This method was applied to the Schwarzschild spacetime \cite{ Chodos:1980df} to find the static electrically charged Einstein-Mawell-dilaton black hole and  to the Kerr spacetime \cite{Frolov:1987rj,Horne:1992zy} to find the corresponding rotating charged dilatonic black hole. In \cite{Rasheed:1995zv} a rotating dyonic black hole in Kaluza-Klein theory was presented. There two boosts and a rotation were applied to the solution to obtain a rotating electrically and magnetically charged black hole in Einstein-Mawell-dilaton theory.

Also in higher dimensions rotating Einstein-Mawell-dilaton black holes were found \cite{Kunz:2006jd} by boosting the higher-dimensional Myers-Perry solution.

In five dimensions Einstein-Mawell-dilaton black rings were constructed. The static charged dilatonic black ring was obtained for arbitrary values of the dilaton coupling constant \cite{Kunduri:2004da} and additionally a rotating charged dilatonic black ring was found in the case of Kaluza-Klein coupling  \cite{Kunduri:2004da, Rocha:2013qya}. In the Kaluza-Klein case also general black ring solutions were constructed: A three-charge black ring \cite{Elvang:2003mj} and a black ring with two angular momenta, electric charge and magnetic charge \cite{Rocha:2012vs, Feldman:2012vd}. Furthermore dilatonic solutions with dipole charges have been obtained, namely a dipole black ring \cite{Yazadjiev:2006ew}, a black saturn with a dipole ring \cite{Yazadjiev:2007cd}, and a solution describing two concentric rotating dipole black rings \cite{Yazadjiev:2008pt}.
Numerically also a seven-dimensional dilatonic black ring was found \cite{Armas:2014rva}.\\

In this paper first a brief review of the rotating charged dilatonic black ring will be given and then the charged dilatonic black saturn solution will be constructed, using the method of boosting the seed solution and then performing a dimensional reduction.  The influence of the dilaton on the black ring and black saturn spacetime is studied by analysing the physical properties and the phase diagram. With increasing charge, the dilatonic solutions need less angular momentum and less horizon area to be balanced. Furthermore, the thermodynamic stability is discussed. The isothermal moment of inertia and specific heat at constant angular momentum of the dilatonic black ring and black saturn cannot be positive at the same time. Therefore the charged dilatonic black ring and black saturn are not thermodynamically stable.

\section{Generating charged dilatonic black objects}

Let us first briefly recall the method and the extraction of the physical properties.

To generate a Kaluza-Klein black hole, the metric of a neutral black object is used as a seed metric. Since we want to construct a five-dimensional object, we take the five-dimensional seed metric and embed it in a six-dimensional spacetime with an extra coordinate $U$,
\begin{equation}
 \dd s_6^2 = \dd s_5^2 + \dd U^2 \, .
\end{equation}
The six-dimensional metric is boosted in the $t$-$U$-plane with the matrix
\begin{equation}
 L = \left( 
      \begin{array}{cc}
       \cosh (\beta ) & \sinh (\beta )\\
       \sinh (\beta ) & \cosh (\beta )
      \end{array}
     \right) \, .
\end{equation}
This boosted metric is still a solution of the (six-dimensional) vacuum Einstein equations. A five-dimensional solution in the  Einstein-Maxwell-dilaton theory can be obtained by comparing the boosted metric to the Kaluza-Klein parametrization of a six-dimensional metric, which reads
\begin{equation}
 \dd s_{6,\rm KK}^2 = \mathrm{e}^{2\alpha\Phi} g_{\mu\nu}\dd x^\mu\dd x^\nu + \mathrm{e}^{-6\alpha\Phi} \left( \dd U + A_\mu \dd x^\mu \right) ^2 \, ,
\end{equation}
where $\alpha=\frac{1}{2\sqrt{6}}$. From the comparison of the transformed $\dd s_6^2$ and  $\dd s_{6,\rm KK}^2$ one can read off the metric components $g_{\mu\nu}$, the Maxwell potential $A_\mu$ and the dilaton function $\Phi$ corresponding to the new five-dimensional charged dilatonic black hole.

The action in five-dimensional Einstein-Maxwell-dilaton theory is
\begin{equation}
 S = \frac{1}{16\pi G}\int \dd^5 x \sqrt{-g} \left( R-\half \Phi_{,\mu}\Phi^{,\mu} -\frac{1}{4} \mathrm{e}^{-2h\Phi} F_{\mu\nu} F^{\mu\nu} \right) \, ,
\end{equation}
where the dilaton coupling constant has the Kaluza-Klein value $h=\frac{2}{\sqrt{6}}$. The obtained charged dilatonic metric fulfils the Einstein equations,
\begin{equation}
 G_{\mu\nu}=8\pi G T_{\mu\nu}
\end{equation}
with
\begin{equation}
 T_{\mu\nu} = \partial _\mu \Phi \partial _\nu \Phi -\half g_{\mu\nu} \partial _\lambda \Phi \partial ^\lambda \Phi + \mathrm{e}^{-2h\Phi} \left( F_{\mu\lambda} F_\nu^\lambda -\frac{1}{4}g_{\mu\nu} F_{\lambda\sigma} F^{\lambda\sigma} \right) \ ,
\end{equation}
the Maxwell equations,
\begin{equation}
 \nabla _\mu \left( \mathrm{e}^{-2h\Phi} F^{\mu\nu} \right) =0 \, ,
\end{equation}
and the dilaton equation,
\begin{equation}
 \nabla ^2 \Phi = -\frac{h}{2}\mathrm{e}^{-2h\Phi} F_{\mu\nu} F^{\mu\nu}\, .
\end{equation}\\

Regarding the physical properties, one can easily find relations between the physical quantities of the charged dilatonic black hole and its uncharged seed metric. In the following the physical quantities of the dilatonic black hole will be denoted with an index $\beta$ and the quantities of the seed metric with an index~$0$.

The (ADM) mass $M$, the angular momentum $J$, the electric charge $Q$, the magnetic moment $\mathcal{M}$, and the dilaton charge $\Sigma$ can be read off from the metric functions at infinity. In spheroidal coordinates $(r,\theta)$ the asymptotic expansion ($r\rightarrow\infty$) of the metric functions yields (\cite{Kunz:2006jd}, \cite{Harmark:2004rm})
\begin{align}
 g_{tt} &\approx -1 + \frac{8GM}{3\pi}\frac{1}{r^2} \label{eqn:mass} \, ,\\
 g_{t\psi} &\approx -\frac{4GJ}{\pi}\frac{\sin(\theta )^2}{r^2} \label{eqn:angmom} \, ,\\
 A_{t} &\approx \frac{4 G Q}{\pi r^2} \label{eqn:charge} \, ,\\
 A_\psi &\approx -\frac{4G\mathcal{M}}{\pi}\frac{\sin(\theta )^2}{r^2} \, \\
 \Phi &\approx \frac{4G\Sigma}{\pi r^2} \label{eqn:dilatoncharge}  \, ,
\end{align}
assuming the black object is five-dimensional and rotates in $\psi$-direction. The physical quantities can now be found by comparing \eqref{eqn:mass} -- \eqref{eqn:dilatoncharge} to the solution in the limit $r\rightarrow\infty$.

Then the following relations between the quantities of the charged and the uncharged black object can be found:
\begin{align}
 M_\beta &= \left(\frac{2}{3}\sinh(\beta)^2+1 \right)  M_0 \, ,\\
 J_\beta &= \cosh(\beta)  J_0 \, ,\\
 Q&= \frac{2}{3}\sinh(\beta)\cosh(\beta) M_0 \, ,\\
 \mathcal{M}&= \sinh (\beta) J_0 \, \\
 \Sigma &= -\left(\frac{2}{3}\right) ^{3/2}\sinh(\beta)^2 M_0 \, .
\end{align}
Then the charges fulfill the relation \cite{ Rasheed:1995zv, Kunz:2006jd}
\begin{equation}
 \frac{Q^2}{M-\frac{\sqrt{6}}{4}\Sigma}= -\frac{2}{\sqrt{6}}\Sigma \, .
\end{equation}
The area of the horizon is
\begin{equation}
 A_H =  \int \! \sqrt{{\rm det}({\rm d}s_H^2)} \,\dd\psi \dd\phi \dd z
\end{equation}
using the canonical coordinates of \cite{Harmark:2004rm}. ${\rm d}s_H^2$ describes the metric at the horizon. Here one gets the relation
\begin{equation}
 A_H^\beta = \cosh(\beta) A_H^0 \, .
\end{equation}
The Hawking temperature can be calculated via the surface gravity
\begin{equation}
 T_H=\frac{\kappa}{2\pi} \, ,
\end{equation}
where
\begin{equation}
 \kappa^2=\left.-\frac{1}{2}(\nabla_\mu\xi_\nu)(\bigtriangledown^\mu\xi^\nu) \right|_{\rm horizon} \, .
\end{equation}
The Killing vector field is
\begin{equation}
 \xi = \partial _t + ¸\Omega \partial _\psi \, .
\end{equation}
The relation between the horizon angular velocity of the dilatonic black hole and the horizon angular velocity of the seed metric is
\begin{equation}
 \Omega_\beta = \frac{1}{\cosh(\beta)} \Omega_0 \, .
\end{equation}
Considering the Hawking temperature a similar relation exists:
\begin{equation}
 T_H^\beta = \frac{1}{\cosh(\beta)} T_H^0 \, .
\end{equation}
The horizon electrostatic potential
\begin{equation}
 \Psi_{\rm el} = \left. \xi^\mu A_\mu \right|_{\rm horizon} = \frac{\sinh (\beta )}{\cosh (\beta )}
\end{equation}
depends only on the boost parameter $\beta$, for the metrics considered in this paper.

\section{The rotating charged dilatonic black ring}

The rotating charged dilatonic black ring was constructed in \cite{Elvang:2003mj, Kunduri:2004da} and analysed in \cite{Liu:2010dq} within the quasilocal formalism, but so far its thermodynamical stability was not considered.\\

\subsection{The solution}

Taking the neutral rotating black ring metric of \cite{Emparan:2001wn} (or more precisely, in the form presented in \cite{Elvang:2003mj}) as seeds, a charged dilatonic rotating black ring can be constructed with the method described above. The metric in ring coordinates is
\begin{align}
 \dd s^2=&-V_\beta(x,y)^{-2/3}\frac{F(x)}{F(y)}\left[ \dd t + (R\sqrt{\lambda\nu}\cosh (\beta ) )(1+y)\dd\psi \right] ^2 \nonumber\\
         &+V_\beta(x,y)^{1/3}\frac{R^2}{(x-y)^2} \left[ -F(x)\left( G(y)\dd\psi^2 +\frac{F(y)}{G(y)}\dd y^2 \right) + F(y)^2 \left( \frac{1}{G(x)}\dd x^2 + \frac{G(x)}{F(x)}\dd\phi^2 \right) \right] \, ,
\end{align}
where $R$ is a scaling parameter,
\begin{align}
 F(\zeta) &= 1-\lambda \zeta \, ,\\
 G(\zeta) &=(1-\zeta^2)(1-\nu\zeta)
\end{align}
and
\begin{equation}
 V_\beta(x,y) = \cosh (\beta )^2 - \sinh (\beta ) ^2\frac{F(x)}{F(y)} \, .
\end{equation}

The coordinate ranges are $-1\leq x\leq 1$ and $-\infty < y \leq -1$,  $\lambda^{-1} < y <\infty$. The parameters $\nu$ and $\lambda$ vary in $0\leq\nu <\lambda\leq 1$. The static charged dilatonic black ring can be obtained by setting $\nu=0$.\\

The nonvanishing parts of the vector potential are
\begin{align}
 A_t &= \frac{\sinh (\beta ) \cosh (\beta ) (F(y)-F(x))}{\cosh (\beta )^2 F(y) - \sinh (\beta )^2 F(x)} \, ,\\
 A_\psi &= \frac{\sinh(\beta) F(x) R\sqrt{\lambda\nu}(1+y) }{\sinh (\beta )^2 F(x) - \cosh (\beta )^2 F(y)} \, ,
\end{align}
and the dilaton function is
\begin{equation}
 \Phi = -\frac{\sqrt{6}}{3} \ln \left( \cosh ( \beta )^2 -\sinh (\beta )^2 \frac{F(x)}{F(y)} \right) \, .
\end{equation}

\subsection{Physical quantities and phase diagram}
The physical quantities are
\begin{align}
 \Omega &= \frac{1}{R\cosh (\beta )} \sqrt{\frac{\nu}{\lambda (1+\lambda )}}\\ 
 M&= \frac{3\pi R^2}{4G}\frac{\lambda (1+\lambda)}{(1+\nu)}\left( 1+\frac{2}{3}\sinh (\beta )^2 \right) \, , \\
 J&= \cosh (\beta ) \frac{\pi R^3}{2G}\frac{\sqrt{\lambda\nu}(1+\lambda)^{5/2}}{(1+\nu)^2} \, ,\\
 Q&= \frac{\pi R^2}{2G} \sinh (\beta )\cosh (\beta )\frac{\lambda (1+\lambda)}{1+\nu} \, ,\\
 \mathcal{M}&=- \frac{\pi R^3}{2G} \sinh (\beta ) \frac{\sqrt{\lambda\nu}(1+\lambda)^{5/2}}{(1+\nu)^2} \, \\
 \Sigma &= \frac{\pi R^2}{2G} \sqrt{\frac{2}{3}} \sinh (\beta )^2 \frac{\lambda (1+\lambda)}{1+\nu} \, ,\\
 A_H&= 8\pi ^2R^3\cosh (\beta )\frac{\sqrt{\lambda (\lambda -\nu)}(\lambda -\nu)(1+\lambda)}{(1+\nu^2)(1-\nu)} \, ,\\
 T_H&= \frac{1}{\cosh (\beta )}\frac{1-\nu}{4\pi R\sqrt{\lambda (\lambda-\nu)}} \, ,\label{eqn:dbr-temp}\\
 \Psi_{\rm el} &= \frac{\sinh (\beta )}{\cosh (\beta )} \, . 
\end{align}
From \eqref{eqn:dbr-temp} it can be seen that the only extremal ($T_H=0$ ) black ring solution is $\nu =1$. This choice of $\nu$ corresponds to a naked singularity \cite{Elvang:2003mj}.\\

In general the charged dilatonic black ring exhibits a conical singularity, which can be placed either at $x=-1$ or $x=+1$. The excess or deficit angle can be computed in ring coordinates as
\begin{equation}
 \delta = 2\pi - \left| \frac{\Delta\phi \partial_x \sqrt{g_{\phi\phi}}}{\sqrt{g_{xx}}} \right| _{x=\pm 1} \, .
\label{eqn:ring-conical-singularity}
\end{equation}
In the following the conical singularity is chosen to be at $x=+1$, so that $\Delta\phi=2\pi\frac{\sqrt{1+\lambda}}{1+\nu}$ and
\begin{equation}
 \delta = 2\pi \left( 1-\frac{(1-\nu)\sqrt{1+\lambda}}{(1+\nu)\sqrt{1-\lambda}} \right) \, .
\end{equation}
The charge parameter $\beta$ does not influence the conical singularity.
In order for the black ring to be balanced ($\delta=0$), the rotation parameters $\nu$ and $\lambda$ have to fulfil the condition
\begin{equation}
 \lambda=\frac{2\nu}{1+\nu^2} \, .
\end{equation}

The sign of the mechanical moment of inertia $I$ divides the charged dilatonic black ring solution into a fat branch ($I>0$) and a thin branch ($I<0$). Keeping $M$ and $\delta$ constant the mechanical moment of inertia is defined by
\begin{equation}
 \frac{1}{I} = \left. \frac{\partial \Omega}{\partial J} \right| _{M,\delta} = \frac{2G(1-\nu)(1+\nu)^2}{R^4\pi \cosh (\beta) (1+\lambda)^3 (3\lambda\nu+\lambda-\nu)} \, .
\label{eqn:inertia-ring}
\end{equation}
In a fixed $Q$ ensemble equation \eqref{eqn:inertia-ring} gives the same result  as in a fixed $\Psi_{\rm el}$ ensemble. The sign of $I$ is not influenced by the charge parameter $\beta$ and the scaling parameter $R$.

\begin{figure}[h] %H
 \centering
 \subfigure[Sign of $\delta$. The solution contains a conical deficit (blue area) or conical excess (green area). The balanced solutions ($\delta=0$) are on the black curve in between the blue and green area.]{
  \includegraphics[width=6cm]{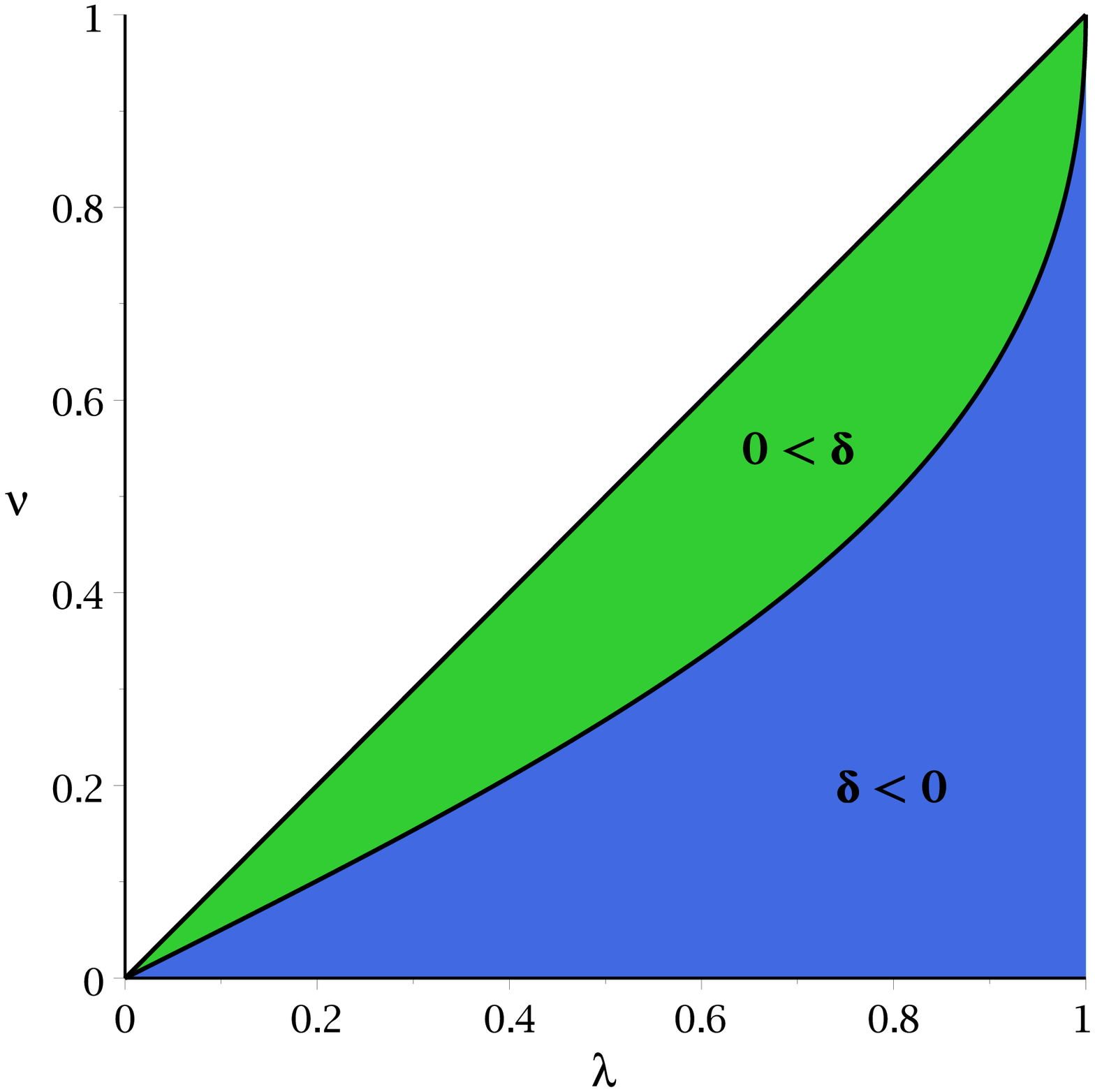}
 }
 \subfigure[Sign of the mechanical moment of inertia. $I>0$ in the green and $I<0$ in the blue area. On the dashed curve the black ring is balanced ($\delta=0$).]{
  \includegraphics[width=6cm]{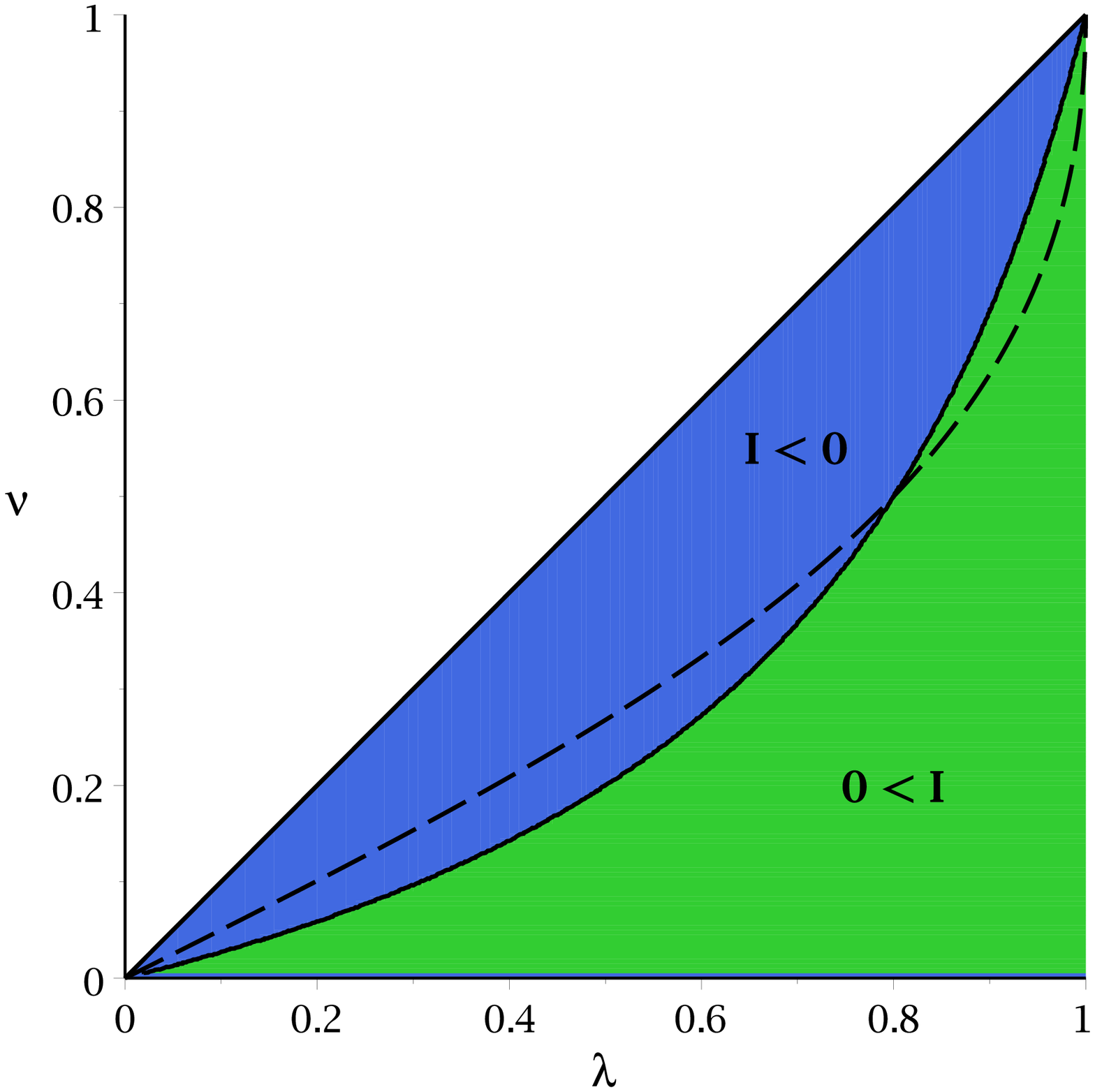}
 }
 \caption{Parameter space of the rotating charged dilatonic black ring}
 \label{pic:ring-delta-I}
\end{figure}

Figure \ref{pic:ring-delta-I} shows the sign of the conical deficit/excess angle (a) and the sign of the mechanical moment of inertia (b) in the $\lambda$-$\nu$-parameter space.\\

For convenience the following scaled quantities are introduced:
\begin{align}
 a_H &= \frac{3}{16}\sqrt{\frac{3}{\pi}} \frac{A_H}{(G M)^{3/2}} \, , \label{eqn:scaled-quantities1}\\
 j^2 &= \frac{27 \pi}{32} \frac{J^2}{G M^3} \, , \\
 q &= \frac{Q}{M} \, , \\
 \tau &= \sqrt{\frac{32\pi}{3}}\sqrt{G M} \ T_H \, . \label{eqn:scaled-quantities2}
\end{align}

\begin{figure}[h]
 \centering
 \subfigure[$a_H$ versus $j^2$ for different values of $\beta$]{
   \includegraphics[width=6cm]{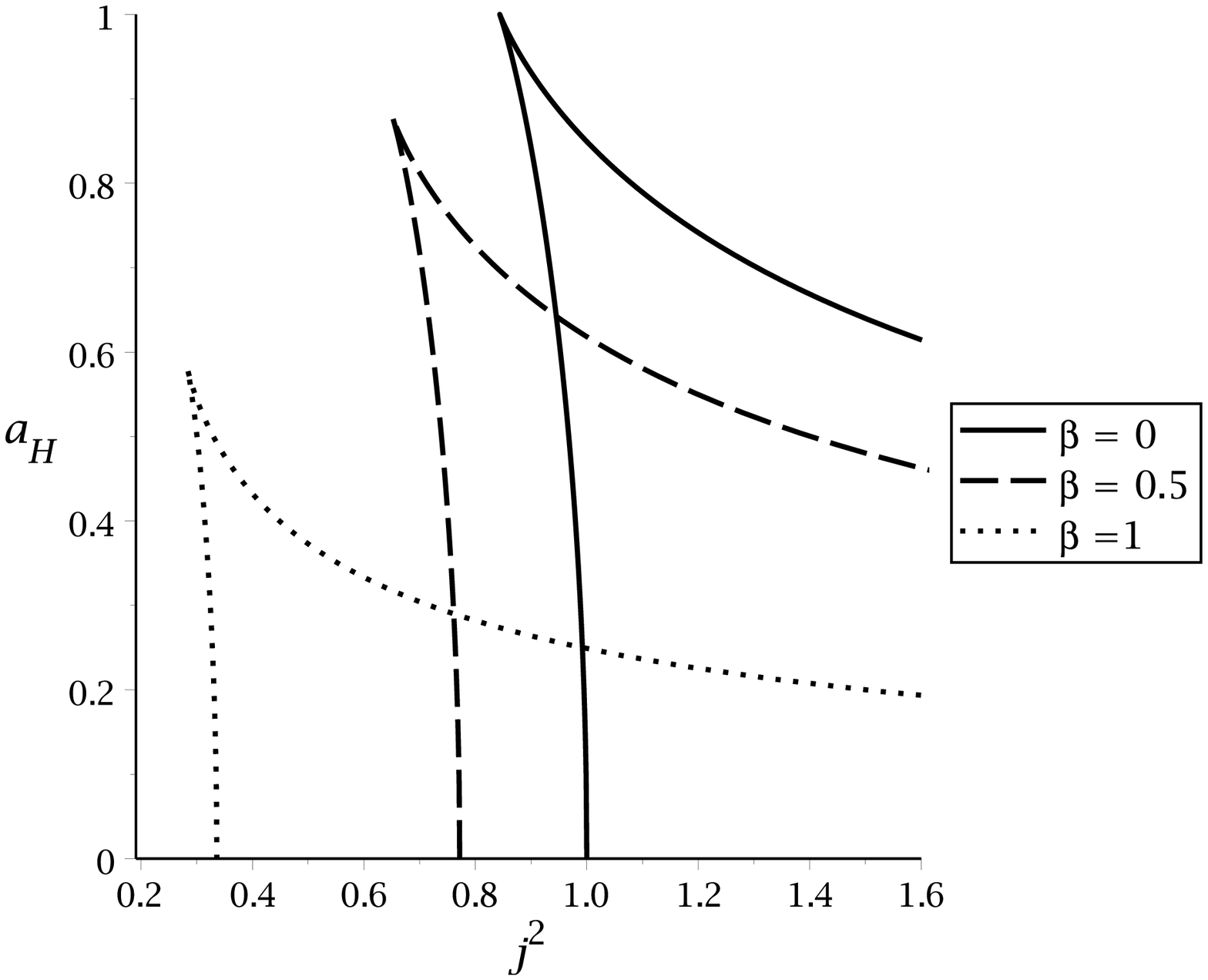}
 }
 \subfigure[$a_H$ versus $j^2$ versus $q$]{
   \includegraphics[width=6cm]{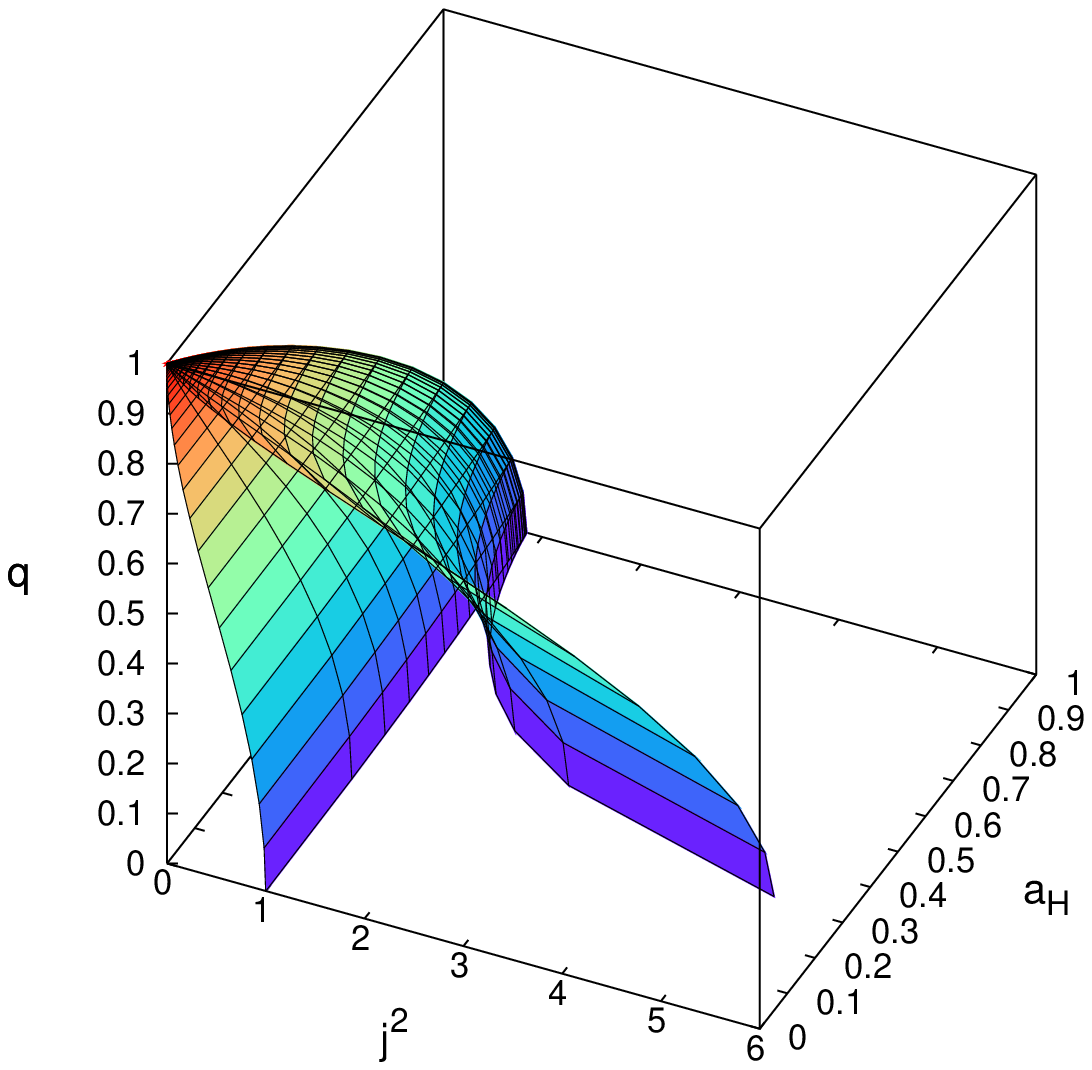}
 }
 \caption{Phase diagram of the charged dilatonic black ring}
 \label{pic:ring_ah-j2-q}
\end{figure}

The scaled charge depends only on $\beta$ and there is an upper limit $q=1$ (and a lower limit $q=-1$) for the charge. The phase diagram of a rotating charged dilatonic black ring is shown in figure \ref{pic:ring_ah-j2-q}, the left picture (a) shows how the $a_H$-$j^2$-diagram changes for different values of $\beta$ and the right picture (b) shows a three-dimensional $a_H$-$j^2$-$q$-diagram. As $\beta$ and accordingly $q$  grows, the curve in the phase diagram gets shifted to lower $a_H$ and $j^2$. The charge comes with an additional force which helps to stabilize the black ring. So when the charge is increased less angular momentum is needed to keep the black ring balanced.\\

\begin{figure}[h]
 \centering
 \subfigure[$j$-$q$-diagram for different values of the parameter $\nu$.]{
  \includegraphics[width=7cm]{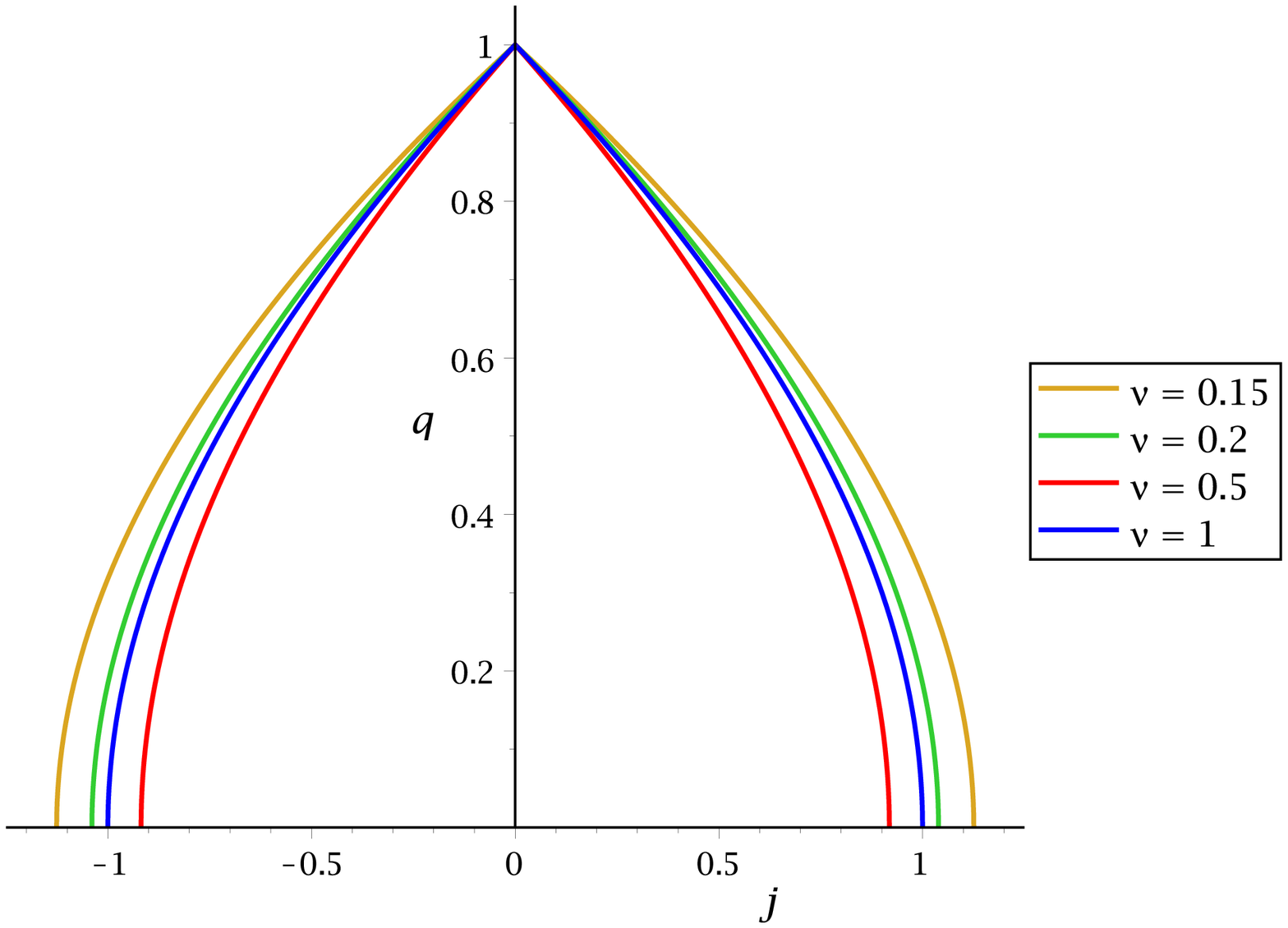}
 }
 \subfigure[$a_H$-$q$-diagram for different values of the parameter $\nu$.]{
  \includegraphics[width=7cm]{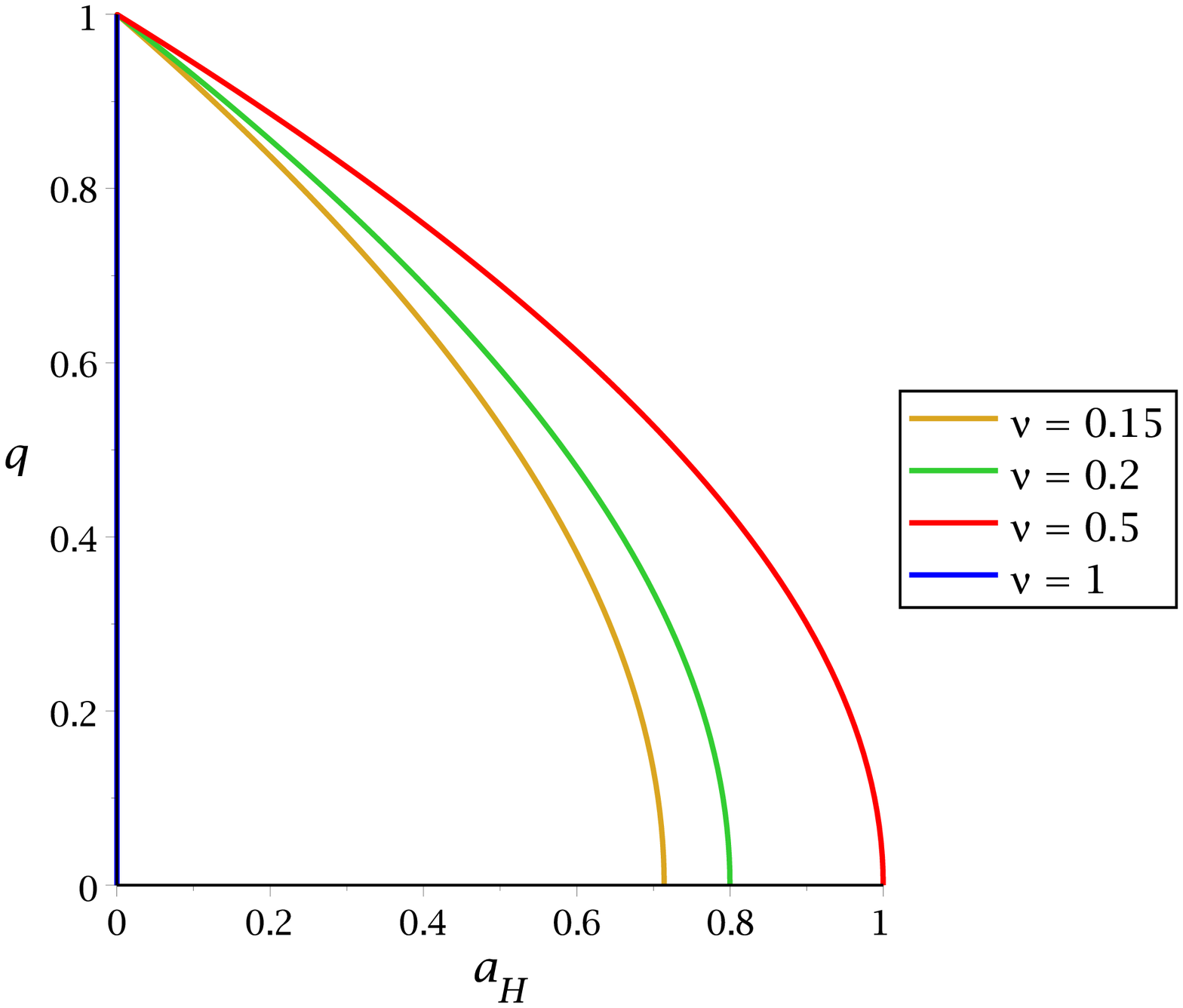}
 }
 \caption{$j$-$q$-diagrams and $a_H$-$q$-diagrams of the charged dilatonic black ring}
 \label{pic:ring_ah-j2-q-2d}
\end{figure}

Figure \ref{pic:ring_ah-j2-q-2d}(a) shows $j$-$q$-plots for different values of the parameter $\nu$, in the extremal case $\nu=1$ the solution is a naked singularity. As for spherical dilatonic black holes \cite{Kunz:2006jd}, the $j$-$q$-plots form a cusp at $j=0$.\\

In figure \ref{pic:ring_ah-j2-q-2d}(b) $a_H$-$q$-diagrams are depicted . The left picture shows diagrams for different values of the parameter $\nu$. $a_H$ reaches its highest value for a given $q$ if $\nu=0.5$, this is also the point where $j^2$ reaches its lowest value. If $\nu=1$, the horizon area vanishes.\\

\begin{figure}[H]
 \centering
 \centering
 \subfigure[Temperature versus $\nu$ for different values of $\beta$.]{
  \includegraphics[width=6cm]{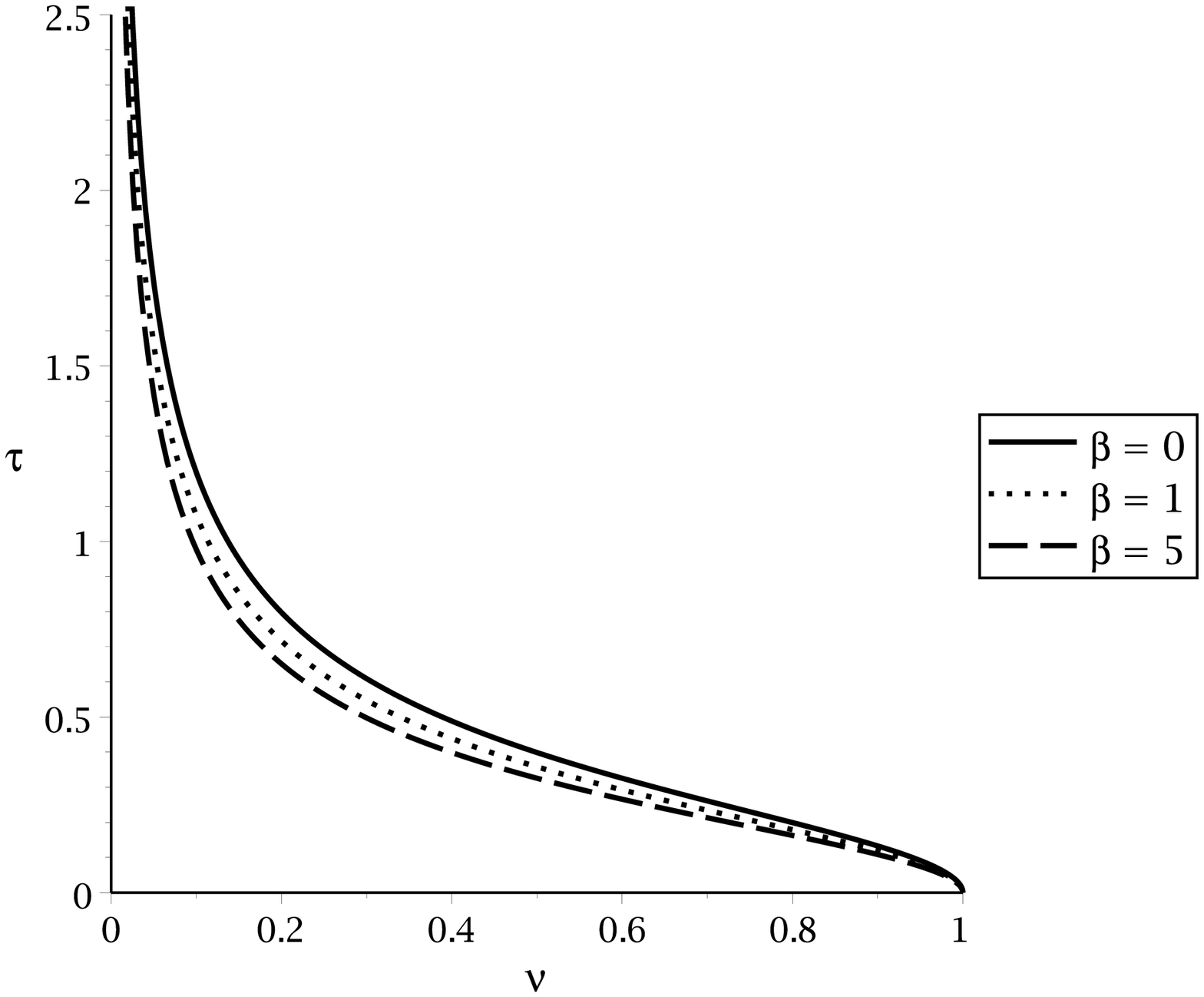}
 }
 \subfigure[Temperature versus $a_H$ versus $q$.]{
  \includegraphics[width=6cm]{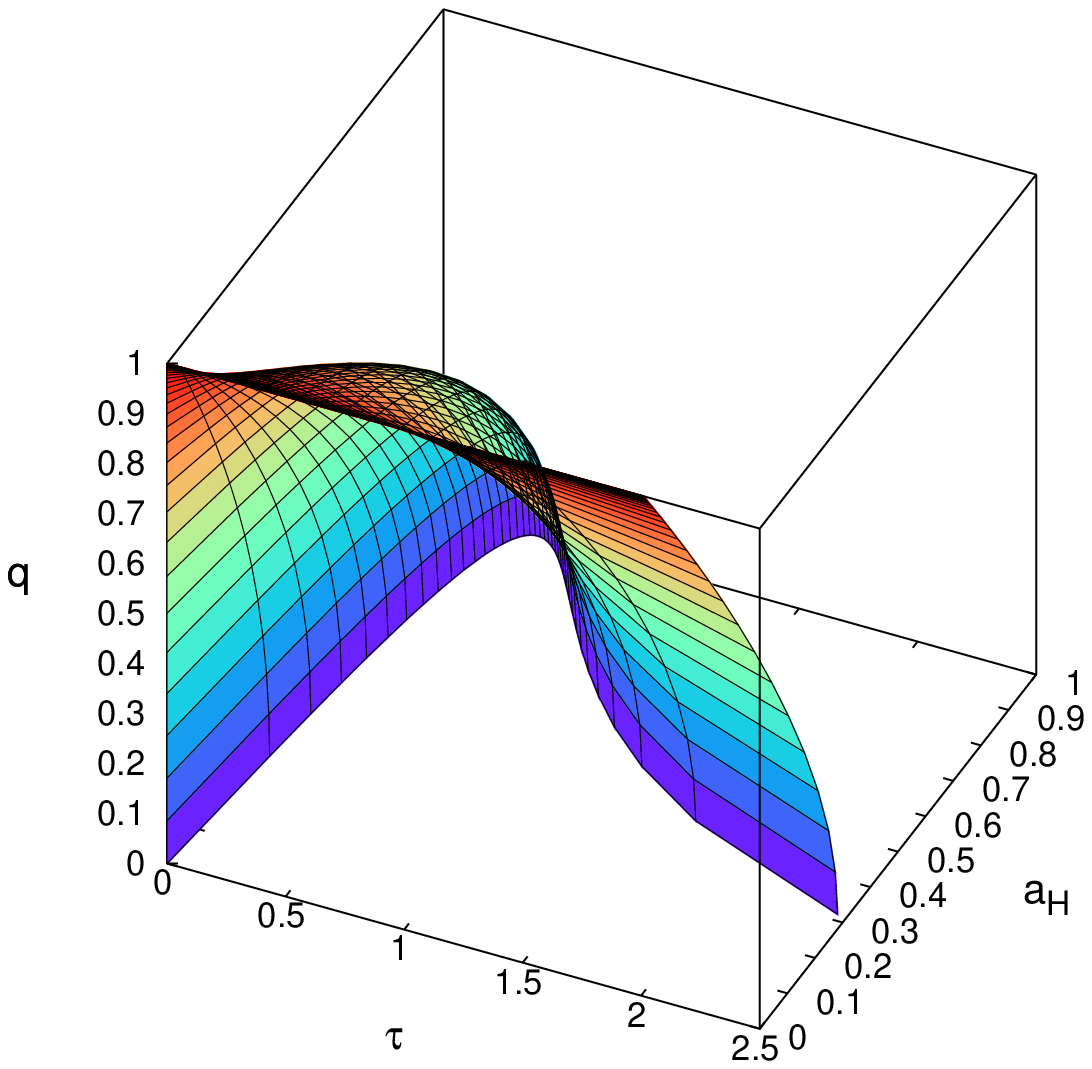}
 }
 \caption{Temperature diagrams of the charged dilatonic black ring}
 \label{pic:ring_temperature}
\end{figure}

The temperature of the charged dilatonic black ring versus the parameter $\nu$ and the temperature versus the horizon area and the charge can be seen in figure \ref{pic:ring_temperature}. The charge has only small influence on the temperature.

\subsection{Thermodynamical stability}
\label{sec:ring-stability}

In a (stationary and asymptotically flat) spacetime that contains a conical singularity, the first law of thermodynamics gets an extra term $ \mathcal{A} \dd P$ (see \cite{Herdeiro:2009vd}, \cite{Herdeiro:2010aq}). $P$ is the pressure caused by the conical singularity and $ \mathcal{A} / T_H$ is the area spanned by the conical singularity.

For the charged dilatonic spacetimes considered in this paper, the first law is
\begin{equation}
 \dd M = T_H \dd S + \Omega \dd J + \Psi_{\rm el} \dd Q - \mathcal{A} \dd P \, .
\end{equation}
The parameter $\mathcal{A}$ for the charged dilatonic black ring is
\begin{equation}
 \mathcal{A}=\pi R^2\frac{\sqrt{1+\lambda} (1-\lambda)^{3/2}}{1-\nu}\, .
\end{equation}
To analyse the thermodynamical stability two quantities are considered (compare \cite{Herdeiro:2010aq}), the specific heat at constant angular momentum
\begin{equation}
 C_J =  \left. T_H \frac{\partial S}{\partial T_H} \right| _{J, \mathcal{A}}
\end{equation}
and the isothermal moment of inertia
\begin{equation}
 \epsilon = \left. \frac{\partial J}{\partial \Omega} \right| _{T_H, \mathcal{A}} \, .
\end{equation}
To ensure thermodynamical stability both quantities need to be positive.\\

In a fixed $Q$ ensemble the specific heat and the isothermal moment of inertia read
\begin{align}
 C_J &= \frac{2 \pi^2 R^3 \cosh(\beta) \sqrt{\lambda} (1+\lambda)(\lambda-\nu)^{3/2} [ 6\cosh(\beta)^2 \lambda (\lambda\nu-\lambda+\nu-1) + \lambda^2(1-3\nu) + 2(\lambda\nu+\lambda-\nu) ] }{G (\nu^2-1) [ 2\cosh(\beta)^2\lambda (\lambda\nu^2+14\lambda\nu-13\nu^2+\lambda-2\nu-1) - \lambda^2(\nu^2-12\nu-3) -12\lambda\nu(1+\nu) + 8\nu^2  ] }\\
 \epsilon &= \frac{\pi R^4 \cosh(\beta)^2 \lambda (1+\lambda)^3}{2G(1+\nu)^3 [ 2\lambda\cosh(\beta)^2(\lambda-1)(\nu-1) + \lambda^2(3-\nu) -2\nu ] }\nonumber\\
          & \times [ 2\lambda\cosh(\beta)^2(-\lambda\nu^2-14\lambda\nu+13\nu^2+2\nu+\lambda+1) + \lambda^2(\nu^2+12\nu+3) -12\lambda\nu(1+\nu) +8\nu^2 ]
\end{align}
and in a fixed $\Psi_{\rm el}$ ensemble the two quantities are
\begin{align}
 C_J &=  \frac{2 \pi^2 R^3 \cosh(\beta) \sqrt{\lambda} (1+\lambda)(\lambda-\nu)^{3/2} [ \lambda^2(3\nu-5) + 2\lambda (4\nu-2) -2\nu ] }{G (\nu^2-1) [ \lambda^2(\nu^2+16\nu-1) +2\lambda(-7\nu^2+4\nu-1) -8\nu^2 ] }\\
 \epsilon &= \frac{\pi R^4 \cosh(\beta)^2 \lambda (1+\lambda)^3 [ \lambda^2(1-16\nu-\nu^2) + 2\lambda(7\nu^2-4\nu+1) + 8\nu^2 ] }{2G(1+\nu)^3 [ \lambda^2(1+\nu) + 2\lambda(1-\nu) -2\nu ] } \, .
\end{align}
In the fixed $\Psi_{\rm el}$ ensemble the sign of $C_J$ and $\epsilon$ is the same for $\beta\neq0$ and $\beta=0$, whereas in the fixed $Q$ ensemble, the charge parameter $\beta$ influences the sign of $C_J$ and $\epsilon$. Nevertheless for all $\beta$ the charged dilatonic black ring is not thermodynamically stable, since $C_J$ and $\epsilon$ cannot be positive at the same time.

\begin{figure}[H]
 \centering
 \subfigure[$\beta=0$]{
  \includegraphics[width=5cm]{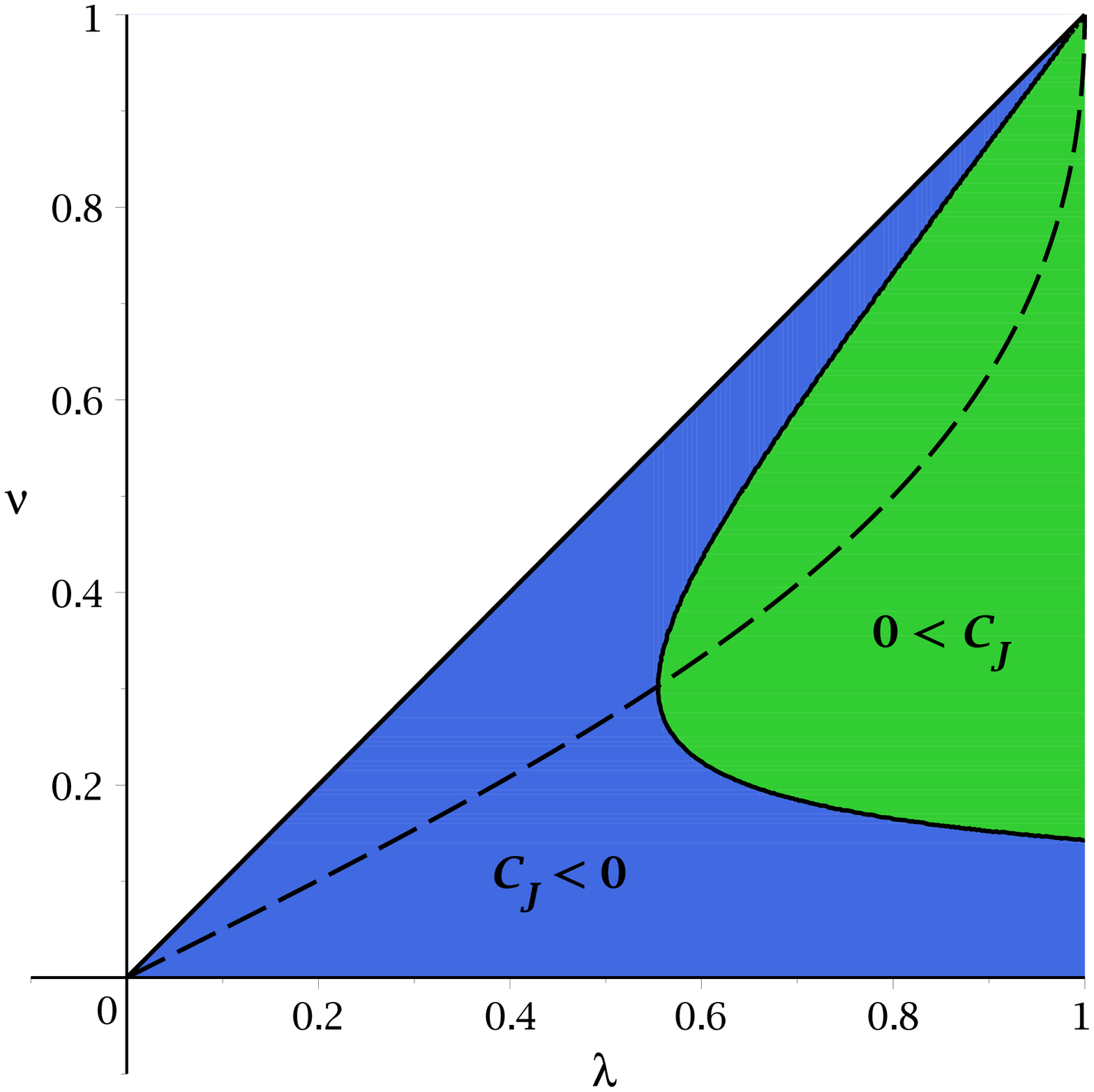}
 }
 \subfigure[$\beta=1$]{
  \includegraphics[width=5cm]{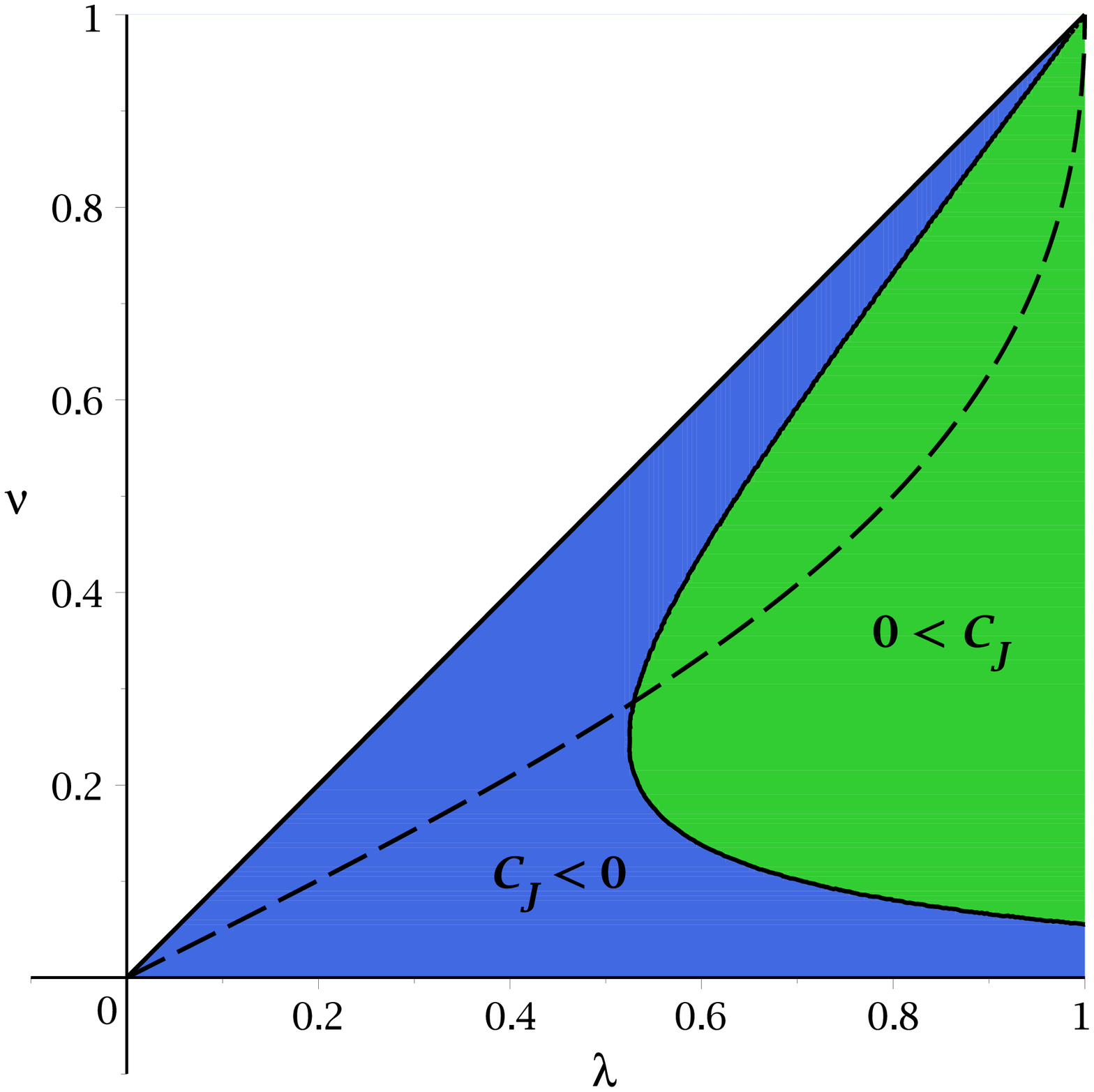}
 }
 \subfigure[$\beta=5$]{
  \includegraphics[width=5cm]{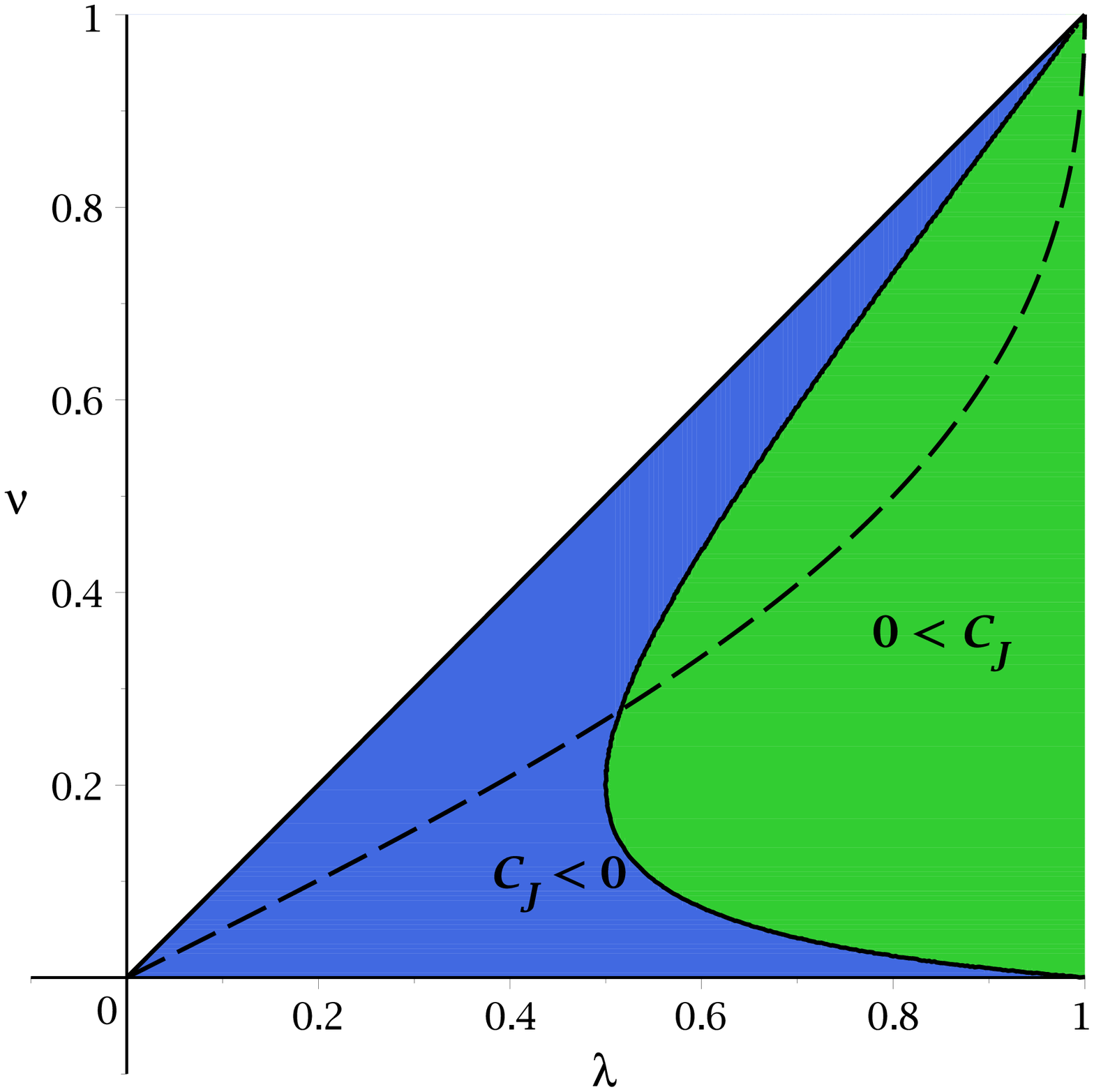}
 }
 \caption{Sign of the specific heat at constant angular momentum of the rotating charged dilatonic black ring in the fixed $Q$ ensemble. $C_J$ is positive in the green area and negative in the blue area. The balanced solutions can be found along the dashed curve.}
 \label{pic:ring_cj}
\end{figure}

\begin{figure}[H]
 \centering
 \subfigure[$\beta=0$]{
  \includegraphics[width=5cm]{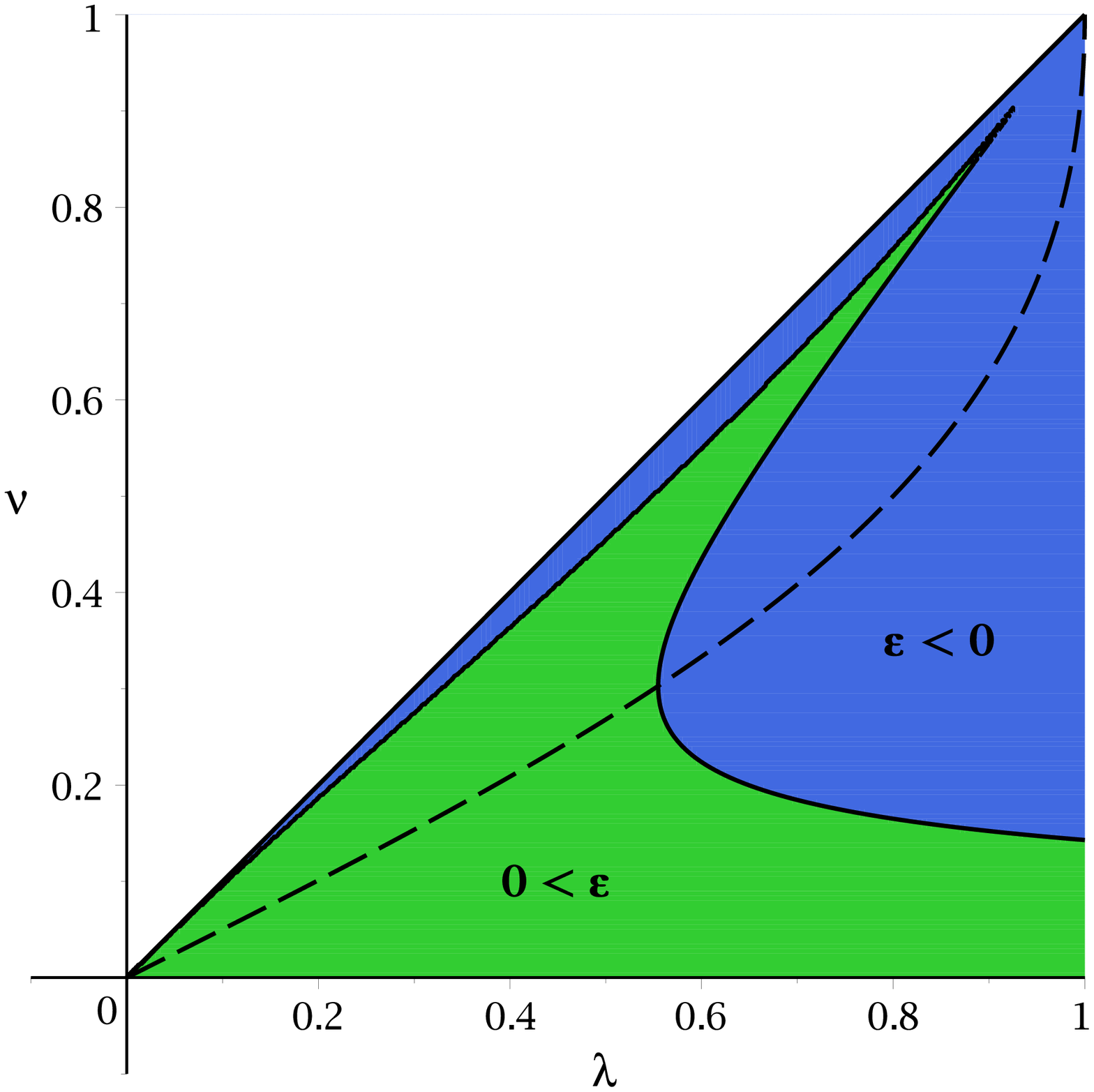}
 }
 \subfigure[$\beta=0.8$]{
  \includegraphics[width=5cm]{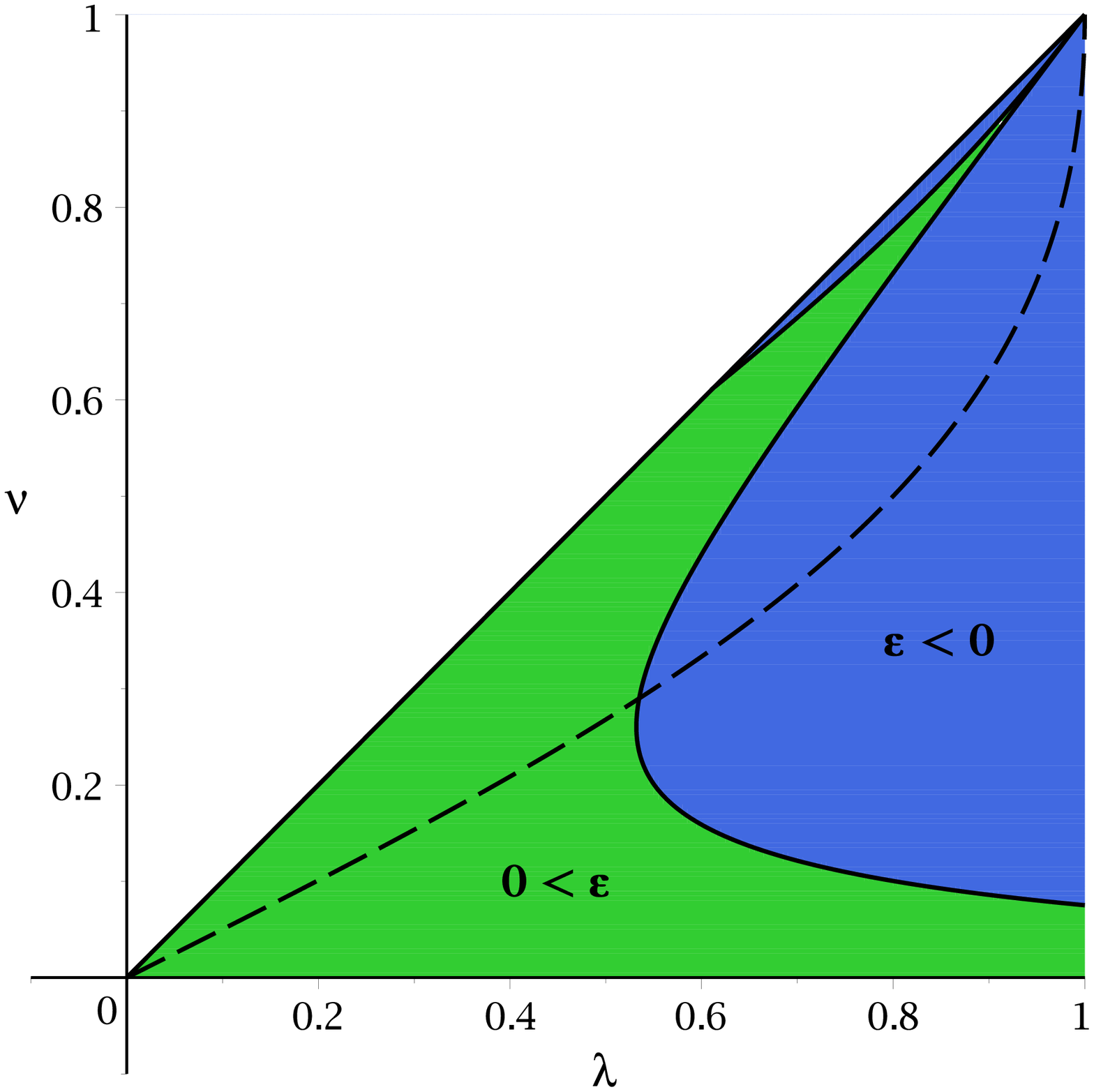}
 }
 \subfigure[$\beta=1.5$]{
  \includegraphics[width=5cm]{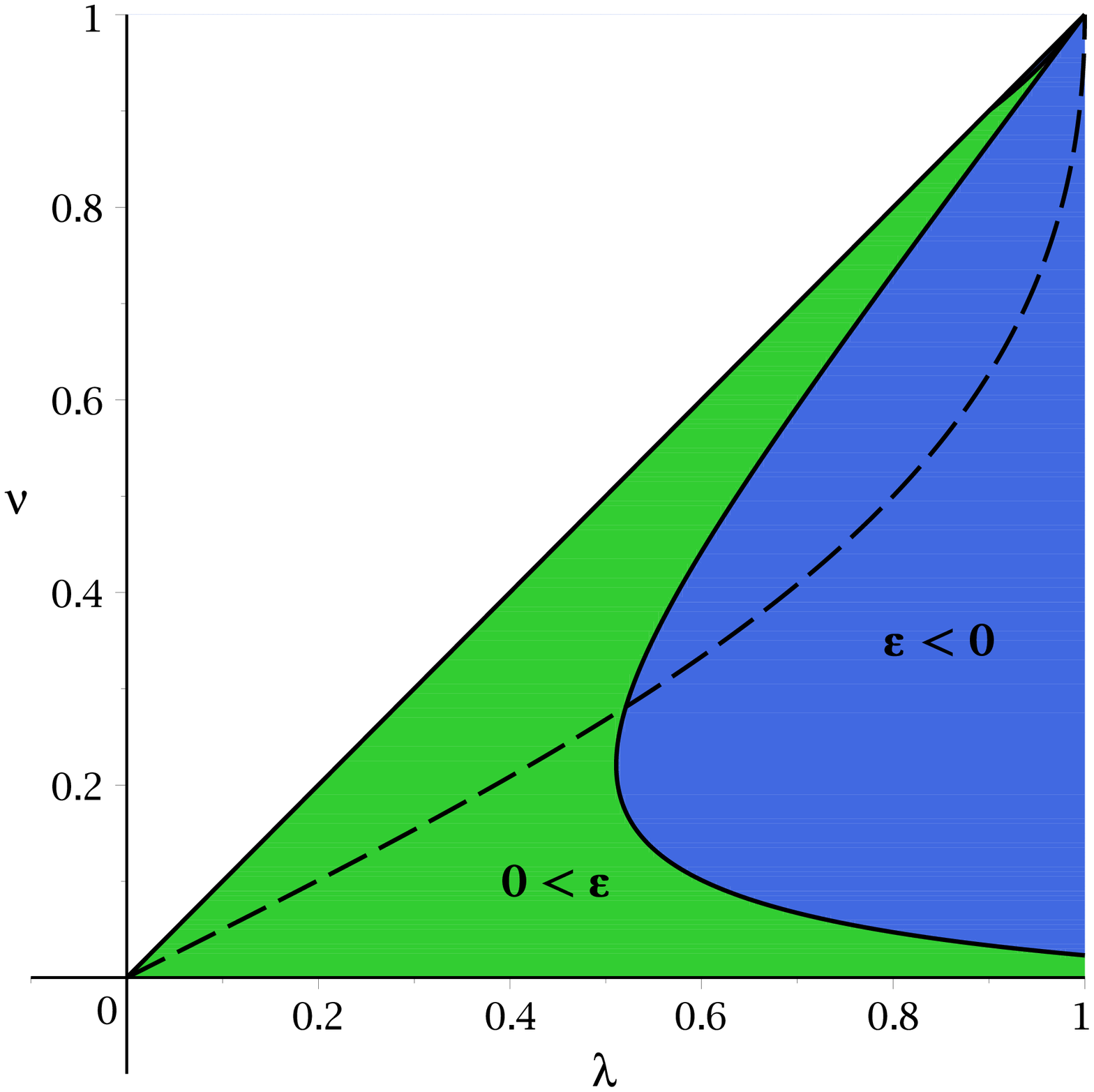}
 }
 \caption{Sign of the isothermal moment of inertia of the rotating charged dilatonic black ring in the fixed $Q$ ensemble. $\epsilon$ is positive in the green area and negative in the blue area. The balanced solutions can be found along the dashed curve.}
 \label{pic:ring_eps}
\end{figure}

Figures \ref{pic:ring_cj} and \ref{pic:ring_eps} show $C_J$ and $\epsilon$ for different values of $\beta$ in the fixed $Q$ ensemble (the plots for $\beta=0$ also correspond to the fixed $\Psi_{\rm el}$ ensemble, since both ensembles are equal at $\beta=0$).

\section{The charged dilatonic black saturn}

\subsection{The solution}
The black saturn solution was found by Elvang and Figueras in 2007 \cite{Elvang:2007rd}. The solution describes a spherical black hole ($S^3$) surrounded by a black ring ($S^1\times S^2$). If one takes the metric of \cite{Elvang:2007rd} as seeds and applies the above methods, one gets the metric of a charged dilatonic black saturn. In canonical coordinates, the metric is given by
\begin{align}
   ds^2 &= - V_\beta (\rho, z)^{-2/3} \, \frac{H_y}{H_x} \left[\dd t + \cosh (\beta ) \left(\frac{\omega_\psi}{H_y}+q\right) \dd\psi \right]^2 \nonumber\\
        &+V_\beta (\rho, z)^{1/3} \, H_x  \left\{ k^2 P \left( \dd\rho^2 + \dd z^2 \right) + \frac{G_y}{H_y}  \dd\psi^2 + \frac{G_x}{H_x}  \dd\phi^2 \right\} \, ,
\end{align}
where
\begin{equation}
 V_\beta (\rho, z) = \cosh (\beta )^2 - \frac{H_y}{H_x} \sinh (\beta )^2 \, .
\end{equation}
The nonvanishing parts of the vector potential are
\begin{align}
 A_t &= \frac{\sinh (\beta ) \cosh (\beta ) (H_x-H_y)}{\cosh (\beta )^2 H_x - \sinh (\beta )^2 H_y} \, ,\\
 A_\psi &= \frac{\sinh(\beta) (\omega_\psi + q H_y)}{\sinh (\beta )^2 H_y - \cosh (\beta )^2 H_x} \,
\end{align}
and the dilaton function is
\begin{equation}
  \Phi = -\frac{\sqrt{6}}{3} \ln \left( \cosh ( \beta )^2 -\sinh (\beta )^2 \frac{H_y}{H_x} \right) \, .
\end{equation}

The metric funtions are
\begin{align*}
 G_x &= \rho^2\frac{\mu_4}{\mu_3 \mu_5} \, , \\
 G_y &= \frac{\mu_3 \mu_5}{\mu_4} \, , \\
 H_x &= F^{-1} \left( M_0 + c_1^2 M_1 + c_2^2 M_2 + c_1 c_2 M_3 + c_1^2 c_2^2 M_4 \right) \, , \\
 H_y &= F^{-1} \frac{\mu_3}{\mu_4} \left( M_0 \frac{\mu_1}{\mu_2} - c_1^2 M_1 \frac{\rho^2}{\mu_1\mu_2} - c_2^2 M_2 \frac{\mu_1\mu_2}{\rho^2} + c_1 c_2 M_3 + c_1^2 c_2^2 M_4 \frac{\mu_2}{\mu_1} \right) \, , \\
 \omega_\psi &= \frac{2}{F \sqrt{G_x}}\left( c_1 R_1\sqrt{M_0 M_1} -c_2 R_2\sqrt{M_0 M_2} + c_1^2c_2 R_2 \sqrt{M_1 M_4} -c_1c_2^2 R_1 \sqrt{M_2 M_4} \right)\, , \\
 P &=  (\mu_3\, \mu_4+ \rho^2)^2 (\mu_1\, \mu_5+ \rho^2) (\mu_4\, \mu_5+ \rho^2) \, \\
 F &= \mu_1 \mu_5  (\mu_1-\mu_3)^2(\mu_2-\mu_4)^2 (\rho^2+\mu_1\mu_3) (\rho^2+\mu_2\mu_3) (\rho^2+\mu_1\mu_4) (\rho^2+\mu_2\mu_4) \\
   &\times  (\rho^2+\mu_2\mu_5) (\rho^2+\mu_3\mu_5) \prod_{i=1}^5 (\rho^2+\mu_i^2) \,
\end{align*}
where
\begin{align*}
 M_0 &= \mu_2 \mu_5^2 (\mu_1-\mu_3)^2 (\mu_2-\mu_4)^2 (\rho^2+\mu_1\mu_2)^2 (\rho^2+\mu_1\mu_4)^2 (\rho^2+\mu_2\mu_3)^2 \, , \\
 M_1 &= \mu_1^2 \mu_2 \mu_3 \mu_4 \mu_5 \rho^2 (\mu_1-\mu_2)^2 (\mu_2-\mu_4)^2(\mu_1-\mu_5)^2 (\rho^2+\mu_2\mu_3)^2  \, , \\
 M_2 &= \mu_2 \mu_3 \mu_4 \mu_5 \rho^2 (\mu_1-\mu_2)^2 (\mu_1-\mu_3)^2 (\rho^2+\mu_1\mu_4)^2(\rho^2+\mu_2 \mu_5)^2  \, ,\\
 M_3 &= 2 \mu_1 \mu_2 \mu_3 \mu_4 \mu_5 (\mu_1-\mu_3) (\mu_1-\mu_5) (\mu_2-\mu_4) (\rho^2+\mu_1^2)(\rho^2+\mu_2^2) (\rho^2+\mu_1\,\mu_4)(\rho^2+\mu_2\, \mu_3) (\rho^2+\mu_2 \mu_5)  \, ,\\
 M_4 &= \mu_1^2 \mu_2 \mu_3^2 \mu_4^2 (\mu_1-\mu_5)^2 (\rho^2+\mu_1\mu_2)^2(\rho^2+\mu_2 \mu_5)^2
\end{align*}
and
\begin{align*}
 \mu_i &= \sqrt{\rho^2+(z-a_i)^2} - (z-a_i) \, , \\ 
 R_i &= \sqrt{\rho^2 + (z-a_i)^2} \, .
\end{align*}
The constants $a_i$ with  $i=1,\ldots,5$ correspond to the rod endpoints of the solution and can be transferred into three dimensionless parameters,
\begin{equation}
 \kappa_i = \frac{a_{i+2}-a_1}{L^2} \, , \quad i=1,\ldots,3
\end{equation}
using a scale parameter $L$ with $L^2 =a_2-a_1 $  \cite{Elvang:2007rd}, where the $\kappa_i$ are ordered in the following way
\begin{equation}
 0 \leq \kappa_3 \leq \kappa_2 < \kappa_1 \leq 1 \, .
\end{equation}
Analogously the coordinate $z$ can be scaled
\begin{equation}
 \overline{z} = \frac{z-a_1}{L^2}
\end{equation}
and a new dimensionless parameter $\overline{c}_2=\frac{c_2}{c_1(1-\kappa_2)}$ can be introduced. The parameters and the coordinate $z$ were scaled in order to simplify the expressions of the physical quantities.

The parameter $c_1$ is fixed at
\begin{equation}
 |c_1|=L\sqrt{\frac{2\kappa_1\kappa_2}{\kappa_3}}
\label{eqn:condition-c1}
\end{equation}
in order to remove singularities at $\rho=0$, $\overline{z}=0$ resulting from the construction of the black saturn in \cite{Elvang:2007rd}.

To ensure the asymptotic flatness of the solution, the parameters $q$ and $k$ have to be chosen as
\begin{align}
 q&=L\sqrt{\frac{2\kappa_1\kappa_2}{\kappa_3}} \frac{\overline{c}_2}{1+\kappa_2\overline{c}_2} \, ,\\
 k&=\frac{1}{|1+\kappa_2\overline{c}_2|} \, .
\end{align}\\

\subsection{Physical quantities and phase diagram}
To calculate the mass, charge, magnetic moment, angular momentum and dilaton charge of the charged dilatonic black saturn the asymptotic coordinates of \cite{Elvang:2007rd} are used:
\begin{align}
 \rho&= \half r^2\sin 2\theta \\
 z&=\half r^2\cos 2\theta  \, .
\end{align}
The physical quantities can be obtained by taking the limit $r\rightarrow\infty$.
The (ADM) mass, charge, magnetic moment, angular momentum and dilaton charge of the charged dilatonic black saturn are
\begin{align}
 M&= \left(\frac{2}{3}\sinh(\beta)^2+1 \right) \frac{3\pi L^2}{4G}\frac{\kappa_3(1-\kappa_1+\kappa_2)-2\kappa_2\kappa_3(\kappa_1-\kappa_2)\overline{c}_2+\kappa_2\left[\kappa_1-\kappa_2\kappa_3(1+\kappa_1-\kappa_2)\right] \overline{c}_2^2}{\kappa_3 \left[ 1+ \kappa_2 \overline{c}_2 \right]^2} \, , \\
%%%
 J &= \cosh(\beta) \frac{\pi L^3}{G}\frac{1}{\kappa_3 \left[ 1+ \kappa_2 \overline{c}_2 \right]^3} \sqrt{\frac{\kappa_2}{2\kappa_1\kappa_3}} \left\{ \kappa_3^2 -\overline{c}_2\kappa_3\left[ (\kappa_1-\kappa_2)(1-\kappa_1+\kappa_3)+\kappa_2(1-\kappa_3) \right] \right. \nonumber\\ 
         &+\left. \overline{c}_2^2\kappa_2\kappa_3\left[ (\kappa_1-\kappa_2)(\kappa_1-\kappa_3)+\kappa_1(1+\kappa_1-\kappa_2-\kappa_3) \right] -\overline{c}_2^3\kappa_1\kappa_2\left[ \kappa_1-\kappa_2\kappa_3(2+\kappa_1-\kappa_2-\kappa_3)\right] \right\} \, ,\\
%%%
 Q&=\frac{\pi L^2}{2G}\sinh(\beta)\cosh(\beta) \frac{\kappa_3(1-\kappa_1+\kappa_2)-2\kappa_2\kappa_3(\kappa_1-\kappa_2)\overline{c}_2+\kappa_2\left[\kappa_1-\kappa_2\kappa_3(1+\kappa_1-\kappa_2)\right] \overline{c}_2^2}{\kappa_3 \left[ 1+ \kappa_2 \overline{c}_2 \right]^2} \, ,\\
%%%
 \mathcal{M}&= \frac{\pi L^3}{G}\sinh (\beta ) \frac{1}{\kappa_3 \left[ 1+ \kappa_2 \overline{c}_2 \right]^3} \sqrt{\frac{\kappa_2}{2\kappa_1\kappa_3}} \left\{ \kappa_3^2 -\overline{c}_2\kappa_3\left[ (\kappa_1-\kappa_2)(1-\kappa_1+\kappa_3)+\kappa_2(1-\kappa_3) \right] \right. \nonumber\\ 
         &+\left. \overline{c}_2^2\kappa_2\kappa_3\left[ (\kappa_1-\kappa_2)(\kappa_1-\kappa_3)+\kappa_1(1+\kappa_1-\kappa_2-\kappa_3) \right] -\overline{c}_2^3\kappa_1\kappa_2\left[ \kappa_1-\kappa_2\kappa_3(2+\kappa_1-\kappa_2-\kappa_3)\right] \right\} \, ,\\
%%%
 \Sigma &= -\frac{\pi L^2}{2G} \sqrt{\frac{2}{3}}\sinh(\beta)\frac{\kappa_3(1-\kappa_1+\kappa_2)-2\kappa_2\kappa_3(\kappa_1-\kappa_2)\overline{c}_2+\kappa_2\left[\kappa_1-\kappa_2\kappa_3(1+\kappa_1-\kappa_2)\right] \overline{c}_2^2}{\kappa_3 \left[ 1+ \kappa_2 \overline{c}_2 \right]^2} \, .
\end{align}

The horizon area is computed for each horizon separately:
\begin{align}
 A_H^{\rm BH}&= 4 L^3 \pi^2 \cosh(\beta) \sqrt{\frac{2(1-\kappa_1)^3}{(1-\kappa_2)(1-\kappa_3)}} \left( \frac{1 + \frac{\kappa_1 \kappa_2 (1-\kappa_2) (1-\kappa_3)}{\kappa_3(1-\kappa_1)} \overline{c}_2^2}{ \left( 1+\kappa_2 \overline{c}_2 \right)^{2}} \right) \, ,\\
%%%
 A_H^{\rm BR}&= 4 L^3 \pi^2 \cosh(\beta) \sqrt{\frac{2 \kappa_2 (\kappa_2-\kappa_3)^3}{\kappa_1 (\kappa_1-\kappa_3)(1-\kappa_3)}} \left( \frac{1- (\kappa_1-\kappa_2) \overline{c}_2 + \frac{\kappa_1 \kappa_2 (1-\kappa_3)}{\kappa_3} \overline{c}_2^2  }{ \left( 1+\kappa_2 \overline{c}_2 \right)^{2}} \right) \, .
\end{align}
The horizon area of the whole black object is $A_H=A_H^{BH}+A_H^{BR}$. From this one gets the entropy of the charged dilatonic black saturn $S=\frac{A_H}{4G}$.\\

The Hawking temperatures of the spherical black hole and the black ring are
\begin{align}
 T_H^{\rm BH}&= \frac{1}{2 L \pi \cosh(\beta)} \sqrt{\frac{(1-\kappa_2)(1-\kappa_3)}{2(1-\kappa_1)}} \left( \frac{ \left( 1+\kappa_2\, \overline{c}_2 \right)^{2}} {1 + \frac{\kappa_1 \kappa_2 (1-\kappa_2) (1-\kappa_3)}{\kappa_3(1-\kappa_1)} \overline{c}_2^2} \right) \, ,\\
%%%
 T_H^{\rm BR}&= \frac{1}{2 L \pi\cosh(\beta)} \sqrt{\frac{\kappa_1(1-\kappa_3)(\kappa_1-\kappa_3)}{2\kappa_2(\kappa_2-\kappa_3)}} \left( \frac{\left( 1+\kappa_2 \overline{c}_2 \right)^{2}}{1- (\kappa_1-\kappa_2) \overline{c}_2 + \frac{\kappa_1 \kappa_2 (1-\kappa_3)}{\kappa_3} \overline{c}_2^2} \right) \, .
\end{align}

The horizon angular velocities are
\begin{align}
 \Omega_{\rm BH}&=\frac{1}{L\cosh(\beta)} \left( 1+\kappa_2 \overline{c}_2 \right) \sqrt{\frac{\kappa_2 \kappa_3}{2 \kappa_1}} \left( \frac{\kappa_3 (1-\kappa_1) - \kappa_1 (1-\kappa_2) (1-\kappa_3) \overline{c}_2}{\kappa_3(1-\kappa_1) + \kappa_1 \kappa_2 (1-\kappa_2) (1-\kappa_3) \overline{c}_2^2} \right) \, ,\\
%%%
 \Omega_{\rm BR}&= \frac{1}{L\cosh(\beta)} \left( 1+\kappa_2\overline{c}_2 \right) \sqrt{\frac{\kappa_1 \kappa_3}{2 \kappa_2}} \left( \frac{\kappa_3  - \kappa_2 (1-\kappa_3) \overline{c}_2}{\kappa_3 -  \kappa_3 (\kappa_1 -\kappa_2) \overline{c}_2 + \kappa_1 \kappa_2 (1-\kappa_3) \overline{c}_2^2} \right) \, .
\end{align}

It is interesting to see how the total charge is divided between the spherical black hole and the black ring. Each charge can be computed separately using a Gau{\ss} integral
\begin{equation}
 Q_i=\frac{1}{16\pi G}  \int_{H_i} \! \mathrm{e}^{2h\Phi} \ast F \, ,
\end{equation}
where $H_i$ is the black hole or black ring horizon and $F=\dd A$. In terms of the metric components this can be written as
\begin{equation}
 Q_i=\frac{1}{16\pi G}  \int_{H_i} \! \mathrm{e}^{2h\Phi} \frac{g_{\phi\phi} g_{zz}}{\sqrt{-\mathrm{det}g}} \left( g_{t\psi} \partial_\rho A_\psi -  g_{\psi\psi} \partial_\rho A_t \right)  \,\dd\psi \dd\phi \dd z \, .
\end{equation}
So the individual charges of the spherical part and the ring part of the dilatonic black saturn are
\begin{align}
 Q_{BH} &=\frac{\pi L^2}{2G} \frac{\kappa_3 (1-\kappa_1)+\kappa_1\kappa_2 (1-\kappa_2)(1-\kappa_3) \overline{c}_2^2}{\kappa_3 (1+ \overline{c}_2\kappa_2)} \, ,\\
%%%
 Q_{BR} &=\frac{\pi L^2}{2G} \frac{\kappa_2 [1-(1-\kappa_2)\overline{c}_2] [\kappa_3-\kappa_3 (\kappa_1-\kappa_2)\overline{c}_2 +\kappa_1\kappa_2(1-\kappa_3) \overline{c}_2^2 ]}{\kappa_3 (1+ \overline{c}_2\kappa_2)^2}  \, .
\end{align}
The two charges sum up to the total ADM charge $Q$. Additionally there is a relation between the charges $Q_i$ and the Komar masses of the sphere ($M_0^{BH}$) and the ring ($M_0^{BR}$) of the uncharged black saturn (see \cite{Elvang:2007rd} for the Komar masses),
\begin{align}
 Q_{BH} &= \frac{2}{3} \sinh(\beta)\cosh(\beta) M_0^{BH}\, ,\\
%%%
 Q_{BR} &= \frac{2}{3} \sinh(\beta)\cosh(\beta) M_0^{BR}\, .
\end{align}

\subsubsection{Balance and thermodynamical equilibrium}

Next we check the $\phi$- and $\psi$-rod for conical singularities. In canonical coordinates the deficit or excess angle of a conical singularity is (compare equation \eqref{eqn:ring-conical-singularity})
\begin{equation}
 \delta = 2\pi- \Delta\eta \lim _{\rho\rightarrow 0} \frac{\partial_\rho \sqrt{g_{\eta\eta}}}{\sqrt{g_{\rho\rho}}} =  2\pi- \Delta\eta\lim _{\rho\rightarrow 0}\sqrt{\frac{g_{\eta\eta}}{\rho ^2g_{\rho\rho}}} \, ,
\label{eqn:periods}
\end{equation}
where $\eta = \phi, \psi$ and $\Delta\eta$ is the period of the angular coordinate $\phi$ or $\psi$. For a balanced black saturn $\delta =0$ holds, so that the period of an angular coordinate is (see also \cite{Elvang:2007rd}, \cite{Harmark:2004rm})
\begin{equation}
 \Delta\eta =  2\pi \lim _{\rho\rightarrow 0}\sqrt{\frac{\rho ^2g_{\rho\rho}}{g_{\eta\eta}}} \, .
\end{equation}
The charged dilatonic black saturn has a semi-infinite $\psi$-rod at $\overline{z}\in [1,\infty]$ and a semi-infinite $\phi$-rod at $\overline{z}\in [-\infty ,\kappa_3]$. These rods fix the periods to be $\Delta\psi = \Delta\phi = 2\pi$ to ensure asymptotic flatness (for the rod structure of a black saturn see  \cite{Elvang:2007rd}). Using equation \eqref{eqn:periods} on both rods, each time one gets the condition $ |c_1|=L\sqrt{\frac{2\kappa_1\kappa_2}{\kappa_3}}$ which is already known from equation \eqref{eqn:condition-c1}.

If this condition holds, there is still a conical singularity left sitting between the black ring and the $S^3$ black hole in the equatorial plane. By requiring $\Delta\phi = 2\pi$ on the finite $\phi$-rod at $\overline{z}\in [\kappa_2,\kappa_1]$ this conical singularity disappears. Here equation \eqref{eqn:periods} gives
\begin{equation}
 \overline{c}_2 = \frac{1}{\kappa_2}\left[ \frac{\kappa_1-\kappa_2}{\sqrt{\kappa_1} (1-\kappa_2)(1-\kappa_3)(\kappa_1-\kappa_3)} - 1 \right] \, .
\label{eqn:condition-c2}
\end{equation}

The conditions for balanced solutions are the same for the charged dilatonic and the neutral black saturn.\\

\begin{figure}[H] %H
 \centering
 \subfigure[Sign of $\delta$. The solution contains a conical deficit (blue area) or conical excess (green area). The balanced solutions ($\delta=0$) are on the black curve in between the blue and green area.]{
  \includegraphics[width=6cm]{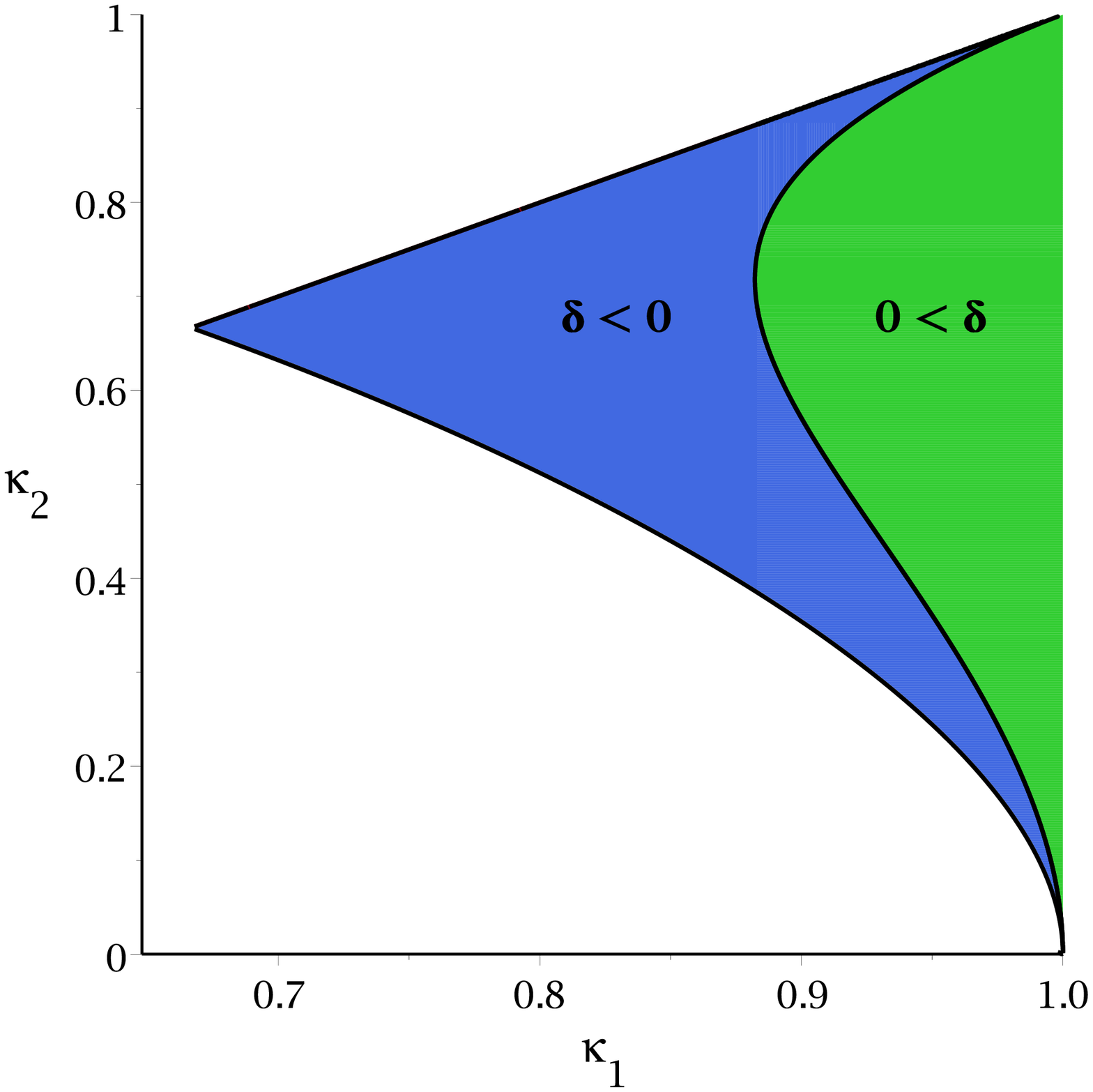}
 }
 \subfigure[Sign of the mechanical moment of inertia. $I>0$ in the green and $I<0$ in then blue area. On the dashed curve the black saturn is balanced ($\delta=0$).]{
  \includegraphics[width=6cm]{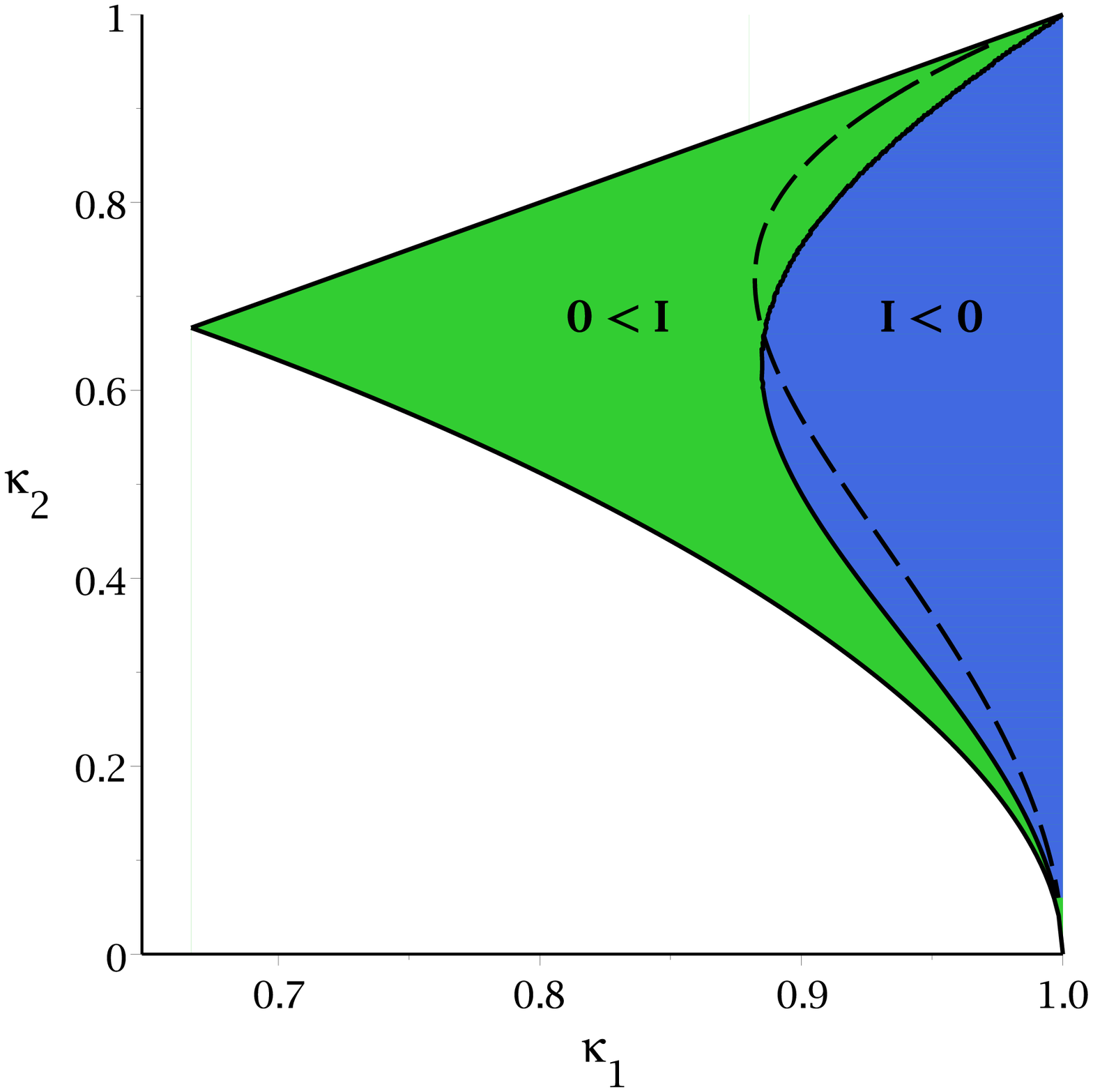}
 }
 \caption{Parameter space of the charged dilatonic black saturn. In the coloured region the solution is in thermodynamical equilibrium.}
 \label{pic:parameterspace-saturn}
\end{figure}

Since the black saturn consists of two black objects, one has to consider under which circumstances the solution is in thermodynamical equilibrium. Clearly, in equilibrium the Hawking temperature of both objects has to be the same. In addition both objects should have the same angular velocity (see \cite{Herdeiro:2010aq}, \cite{Elvang:2007hg}) and the same horizon electrostatic potential. The latter is already fulfilled, so one just needs to consider
\begin{align}
 \Omega_{\rm BH} &= \Omega_{\rm BR} \label{eqn:equilibrium-omega}\, , \\
 T_H^{\rm BH} &= T_H^{\rm BR} \label{eqn:equilibrium-temp} \, .
\end{align}
Equation \eqref{eqn:equilibrium-omega} leads to
\begin{equation}
 \overline{c}_2 = -\frac{\kappa_3 (1-\kappa_1)}{\kappa_1\kappa_2 (1-\kappa_3)}
\end{equation}
and together with \eqref{eqn:equilibrium-temp} one finds
\begin{equation}
 \kappa_3 = \kappa_1 + \frac{\kappa_2 (1-\kappa_2)}{1-\kappa_1-\kappa_2} \, .
\end{equation}
If \eqref{eqn:condition-c2} is not imposed the conical singularity in thermodynamical equilibrium is
\begin{equation}
 \delta = 2\pi \left( 1-\frac{\kappa_2 (1-\kappa_2 )^2}{(\kappa_1-\kappa_2)(\kappa_1+\kappa_2-1)} \sqrt{\frac{\kappa_2}{\kappa_1 [\kappa_1 \kappa_2 (\kappa_1+\kappa_2-1)^2]}} \right) \, .
\end{equation}
Here the relations for the charged dilatonic black saturn and the relations for the neutral black saturn are the same (see \cite{Herdeiro:2010aq}). As in \cite{Herdeiro:2010aq} figure \ref{pic:parameterspace-saturn} shows the parameter space of the charged dilatonic black saturn. In the coloured region the charged dilatonic black saturn is in thermodynamical equilibrium. The sign of $\delta$ can be seen in figure \ref{pic:parameterspace-saturn}(a). On the black curve, the solution is balanced. On the left of the black curve the solution has a conical deficit ($\delta <0$), while on the right side it has a conical excess ($\delta >0$).

The sign of the mechanical moment of inertia is depicted in figure \ref{pic:parameterspace-saturn}(b). $I>0$ corresponds to the fat branch of the solutions and $I<0$ corresponds to the thin branch. Here the balanced solutions can be found along the dashed curve.

If all the conditions for balance and thermodynamical equilibrium are fulfilled at the same time, this leaves one free parameter plus the charge parameter $\beta$.

\subsubsection{Phase diagram}

Here the conditions for balance and thermodynamical equilibrium are imposed, so that $\kappa_2$ and $\beta$ are the only free parameters.

\begin{figure}[H]
 \centering
 \subfigure[$a_H$ versus $j^2$ for different values of $\beta$]{
   \includegraphics[width=6cm]{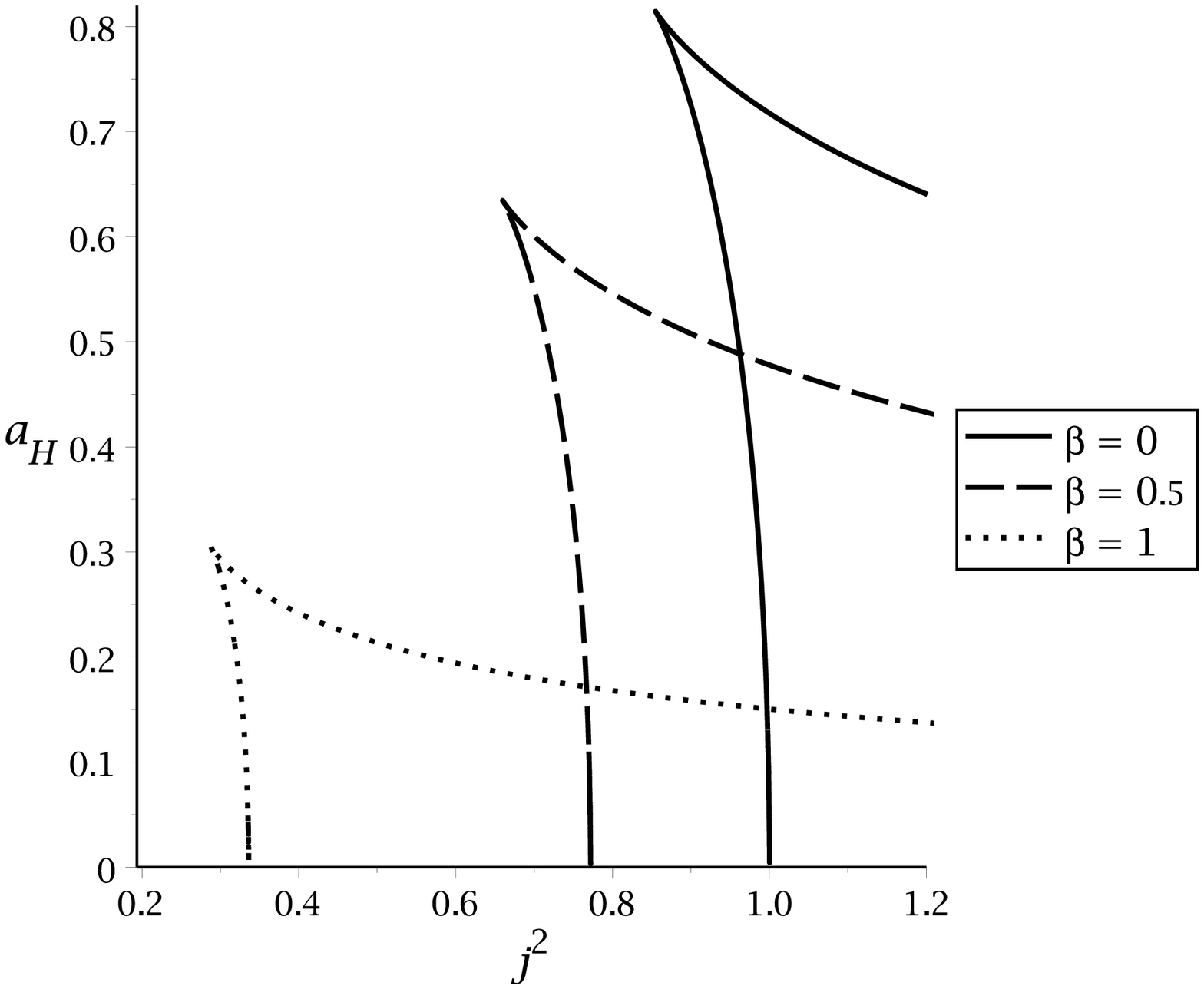}
 }
 \subfigure[$a_H$ versus $j^2$ versus $q$]{
   \includegraphics[width=6cm]{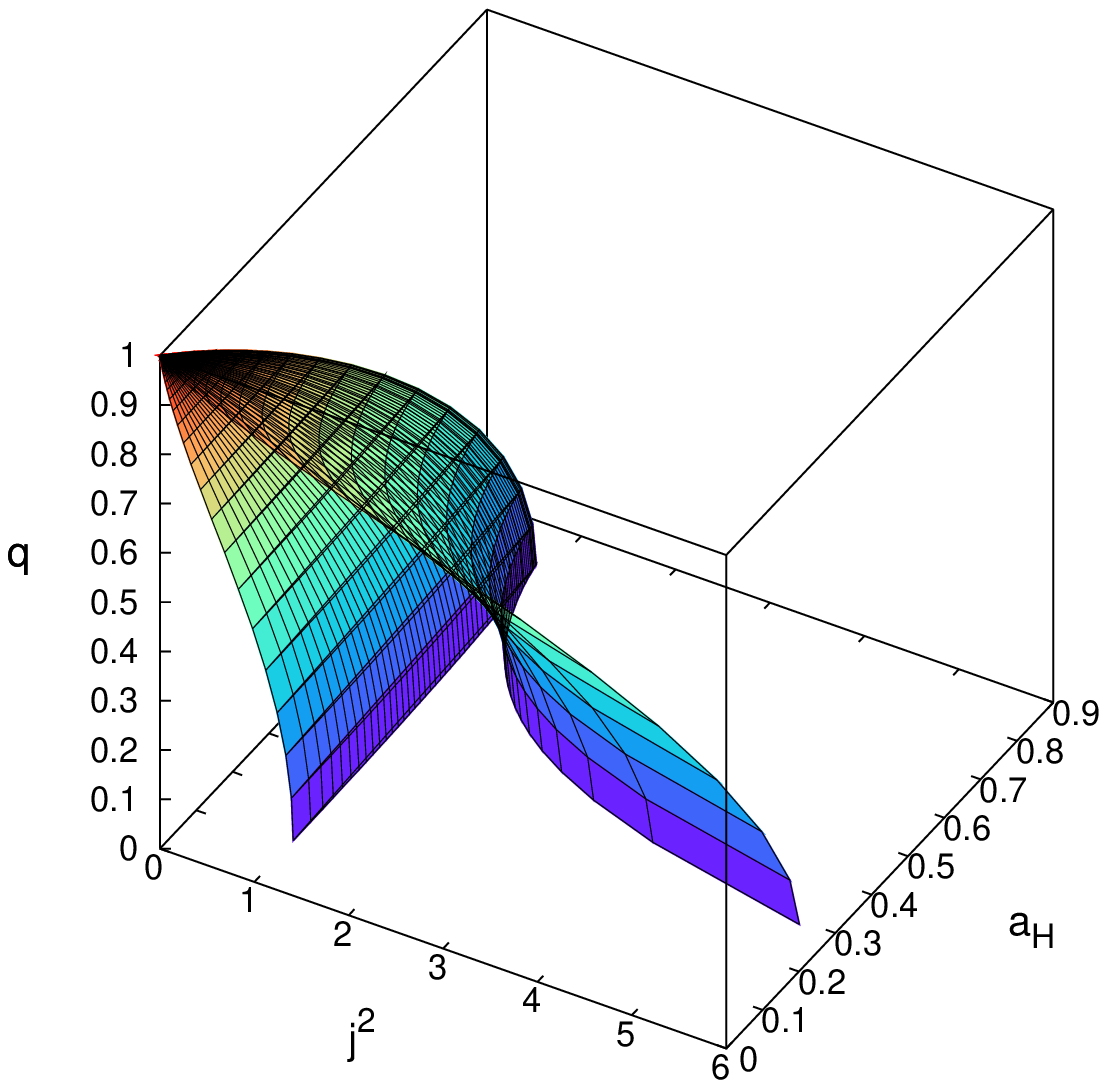}
 }
 \caption{Phase diagram of the charged dilatonic black saturn}
 \label{pic:saturn_ah-j2-q}
\end{figure}

The phase diagram of a rotating charged dilatonic black saturn is shown in figure \ref{pic:saturn_ah-j2-q}, the left picture shows how the $a_H$-$j^2$-diagram changes for different values of $\beta$ and the right picture shows a three-dimensional $a_H$-$j^2$-$q$-diagram. As $\beta$ and accordingly $q$ grows, the curve in the phase diagram gets shifted to lower $a_H$ and $j^2$. The charge comes with an additional force which helps to stabilize the black saturn. So when the charge is increased less angular momentum is needed to keep the black saturn balanced.

As before scaled quantities are used (see equations \eqref{eqn:scaled-quantities1} -- \eqref{eqn:scaled-quantities2}). The scaled charge depends only on $\beta$ and has the upper limit $q=1$ (and the lower limit $q=-1$).\\

\begin{figure}[H]
 \centering
 \subfigure[$j$-$q$-diagram for different values of the parameter $\kappa _2$.]{
  \includegraphics[width=5cm]{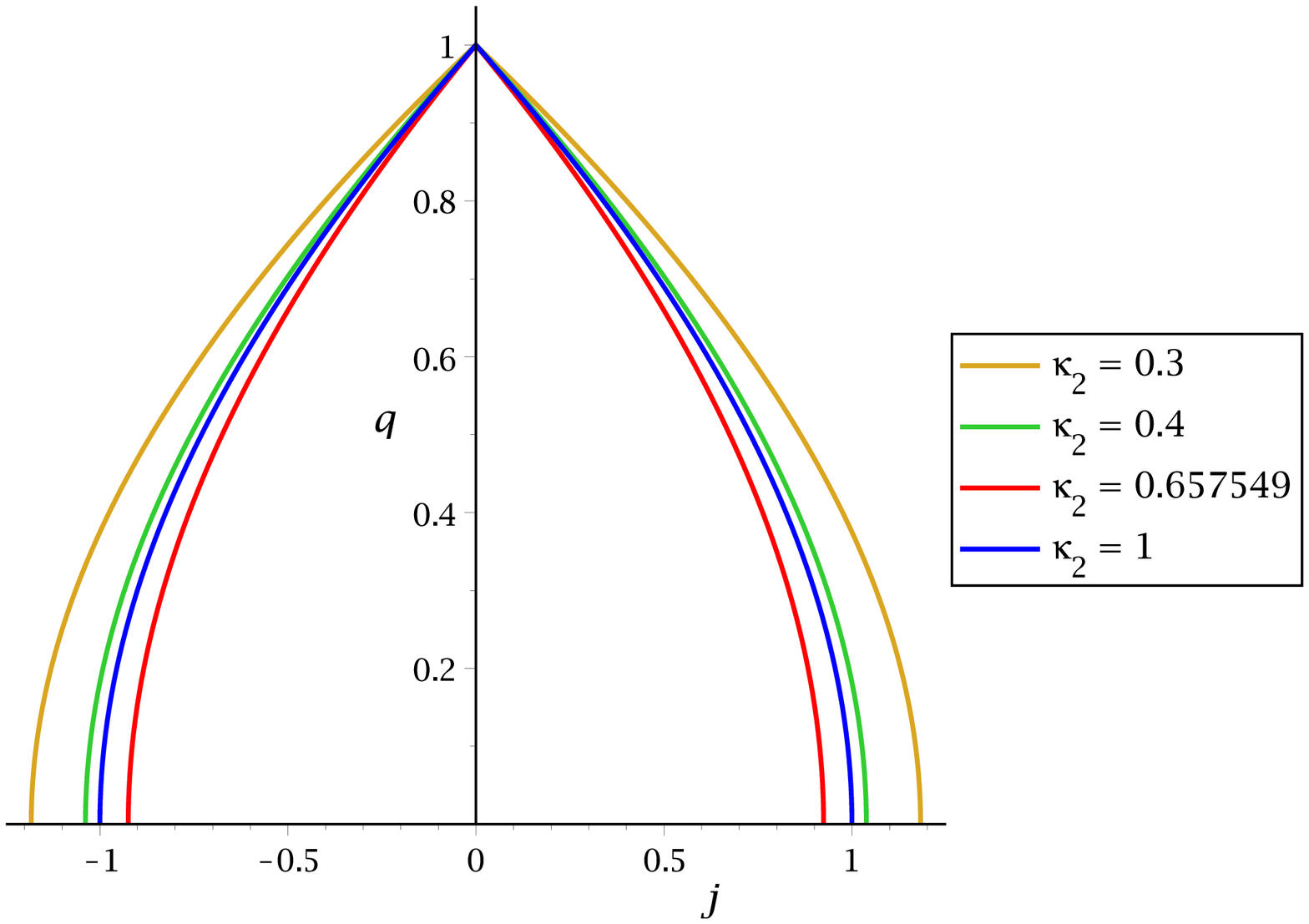}
 }
 \subfigure[$a_H$-$q$-diagram for different values of the parameter $\kappa_2$.]{
  \includegraphics[width=5cm]{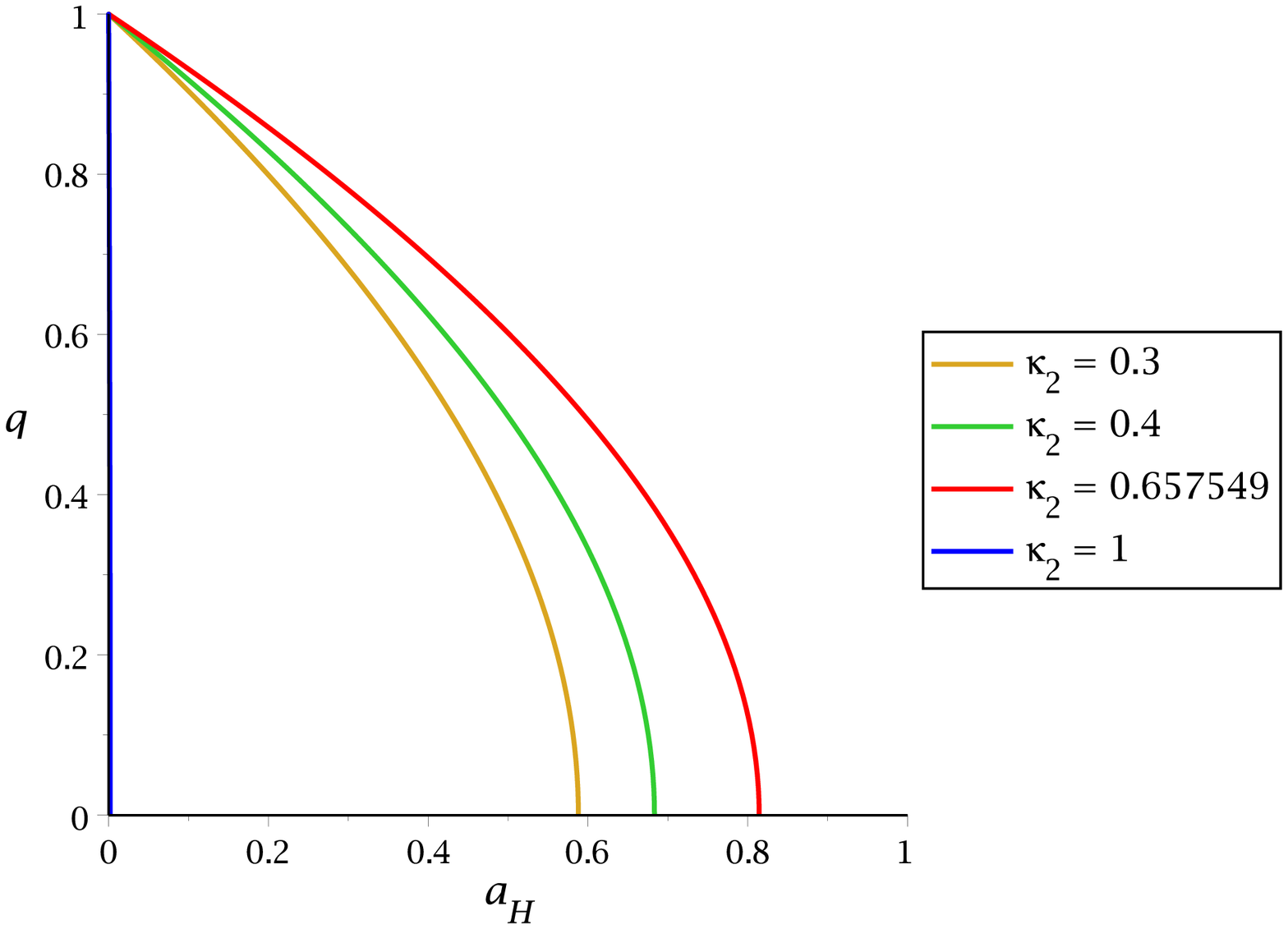}
 }
 \subfigure[Comparison of the $a_H$-$q$-diagrams. Case $\kappa_2\approx 0.657549$ for the charged dilatonic black saturn and $\nu=0.5$ for the charged dilatonic black ring: For a given $q$,  $a_H$ has its maximal value (and $j^2$ its minimal value).]{
  \includegraphics[width=5cm]{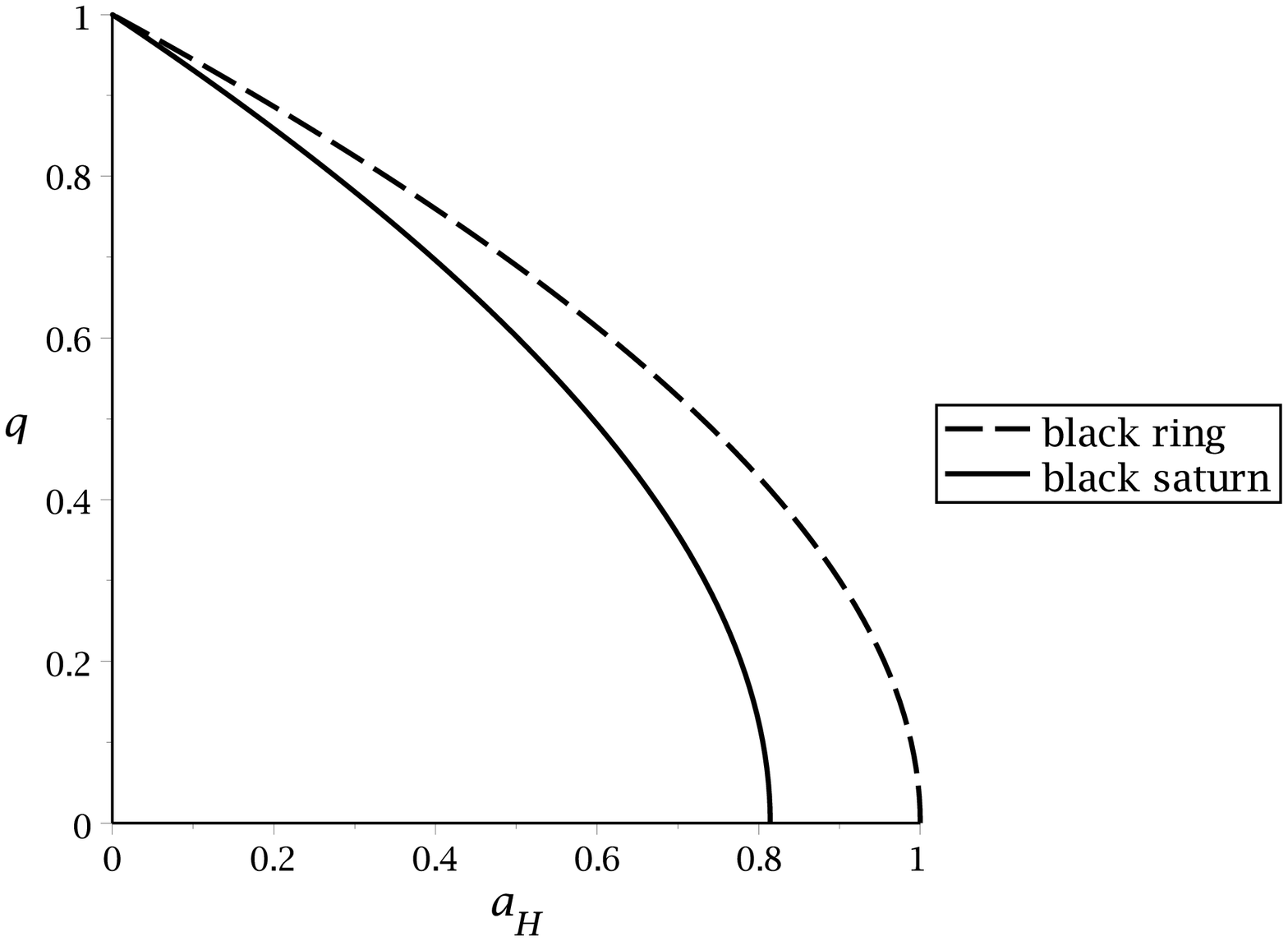}
 }
 \caption{$j$-$q$-diagrams and $a_H$-$q$-diagrams of the charged dilatonic black saturn and a comparison to the charged dilatonic black ring }
 \label{pic:saturn_ah-j2-q-2d}
\end{figure}

Figure \ref{pic:saturn_ah-j2-q-2d}(a) shows $j$-$q$-plots for different values of the parameter $\kappa_2$. In the extremal case $\kappa_2=1$ the solution is a naked singularity. As before the $j$-$q$-plots form a cusp at $j=0$. The $j$-$q$-diagrams of the charged dilatonic black saturn are similar to those of the charged dilatonic black ring, in fact they are exactly the same in the extremal cases $\nu=1$ and $\kappa_2=1$, respectively.\\

$a_H$-$q$-diagrams for different values of the parameter $\kappa_2$ are depicted in figure \ref{pic:saturn_ah-j2-q-2d}(b). $a_H$ reaches its highest value for a given $q$ if $\kappa_2\approx 0.657549$, this is also the point where $j^2$ reaches its lowest value. If $\kappa_2=1$, the horizon area vanishes. \ref{pic:saturn_ah-j2-q-2d}(c) shows a comparison of the $a_H$-$q$-diagrams of the charged dilatonic black ring and the charged dilatonic black saturn, the parameters $\nu$ and $\kappa_2$ are chosen in such a way that $a_H$ is maximal for each $q$.\\

\begin{figure}[H]
 \centering
 \subfigure[Temperature versus $\kappa_2$ for different values of $\beta$.]{
  \includegraphics[width=6cm]{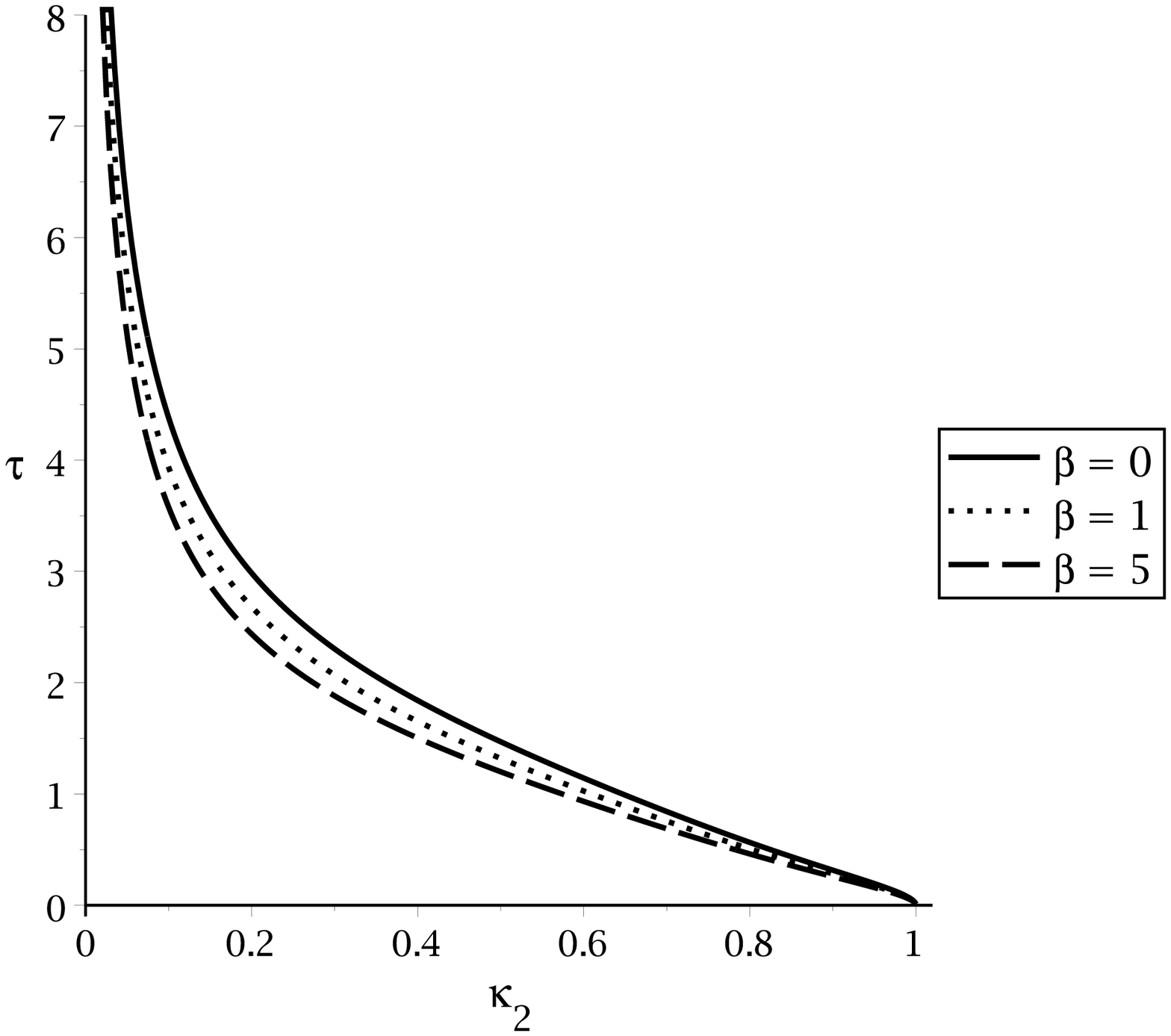}
 }
 \subfigure[Temperature versus $a_H$ versus $q$]{
  \includegraphics[width=6cm]{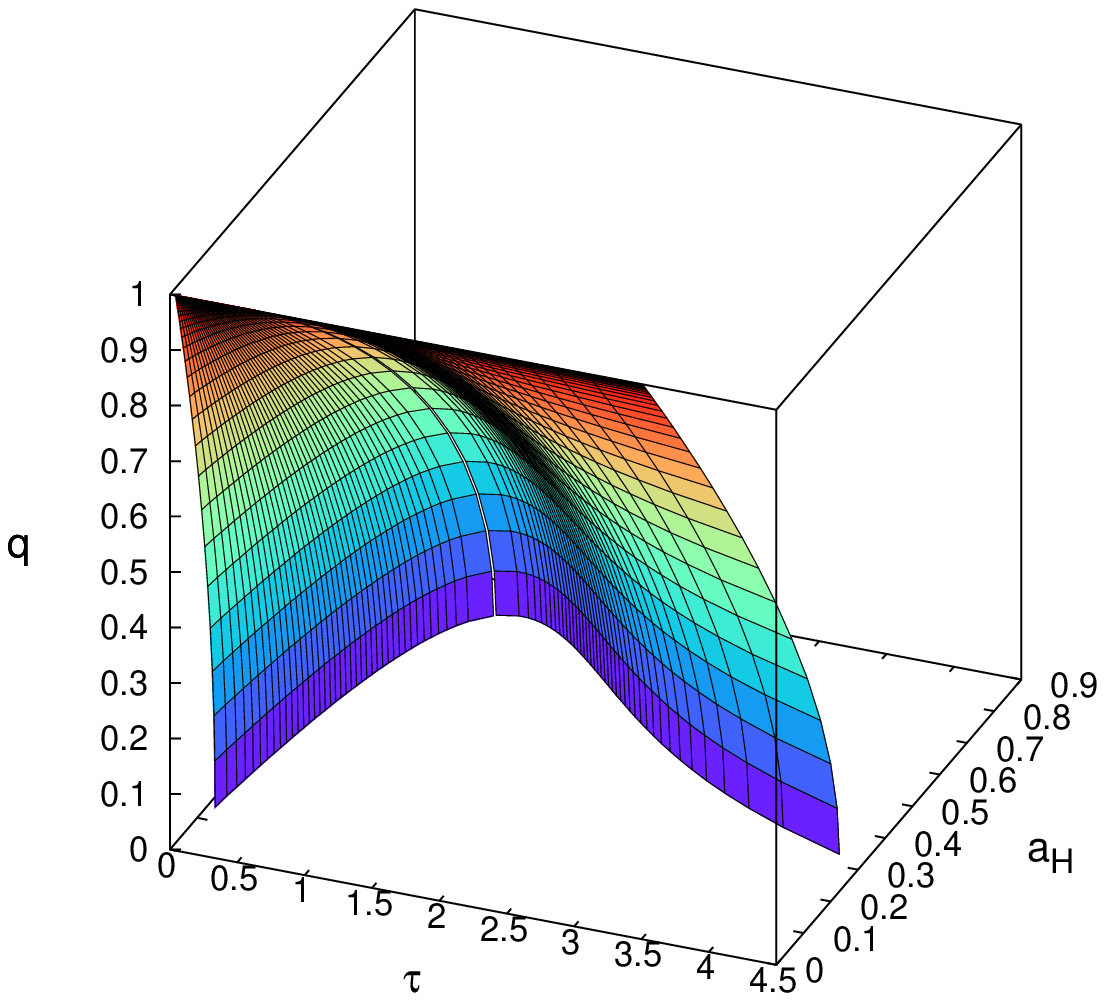}
 }
 \caption{Temperature-diagrams of charged the dilatonic black saturn}
 \label{pic:saturn_temperature}
\end{figure}

The temperature of the charged dilatonic black saturn versus the parameter $\kappa_2$ and the temperature versus the horizon area and the charge can be seen in figure \ref{pic:saturn_temperature}. The charge has only small influence on the temperature.

\subsection{Thermodynamical stability}

To determine the thermodynamical stability of the charged dilatonic black saturn, the specific heat at constant angular momentum $C_J$ and the isothermal moment of inertia $\epsilon$ are calulated as discussed in section \ref{sec:ring-stability}. Due to their enormous size, the exact formulas are not displayed here.\\

The results are similar to those of the  charged dilatonic black ring. In the fixed $\Psi_{\rm el}$ ensemble the sign of $C_J$ and $\epsilon$ is the same for $\beta\neq0$ and $\beta=0$, whereas in the fixed $Q$ ensemble, the charge parameter $\beta$ influences the sign of $C_J$ and $\epsilon$. As in the black ring case for all $\beta$ the charged dilatonic black saturn is not thermodynamically stable, since $C_J$ and $\epsilon$ cannot be positive at the same time.

\begin{figure}[H]
 \centering
 \subfigure[$\beta=0$]{
  \includegraphics[width=5cm]{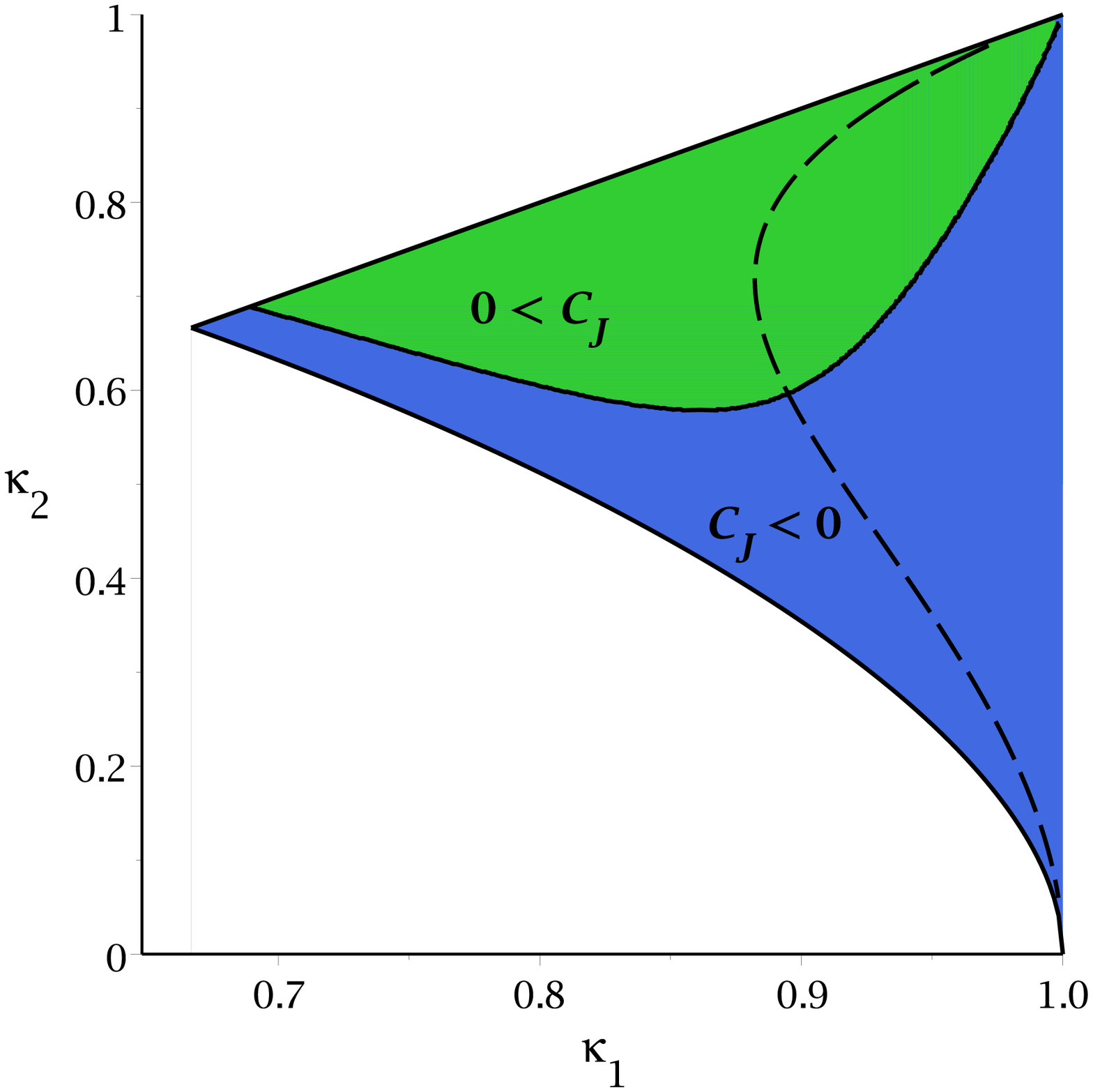}
 }
 \subfigure[$\beta=1$]{
  \includegraphics[width=5cm]{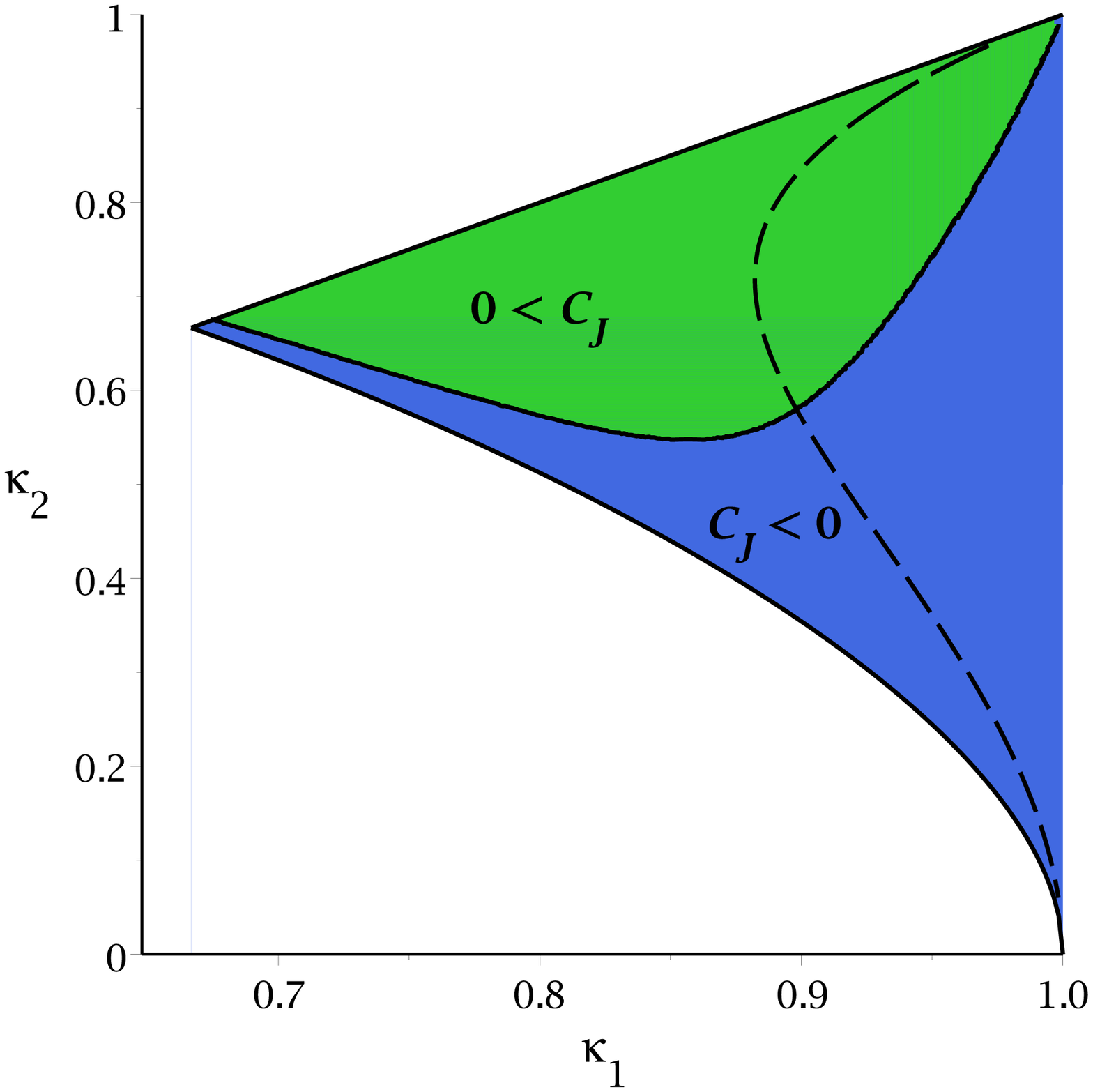}
 }
 \subfigure[$\beta=5$]{
  \includegraphics[width=5cm]{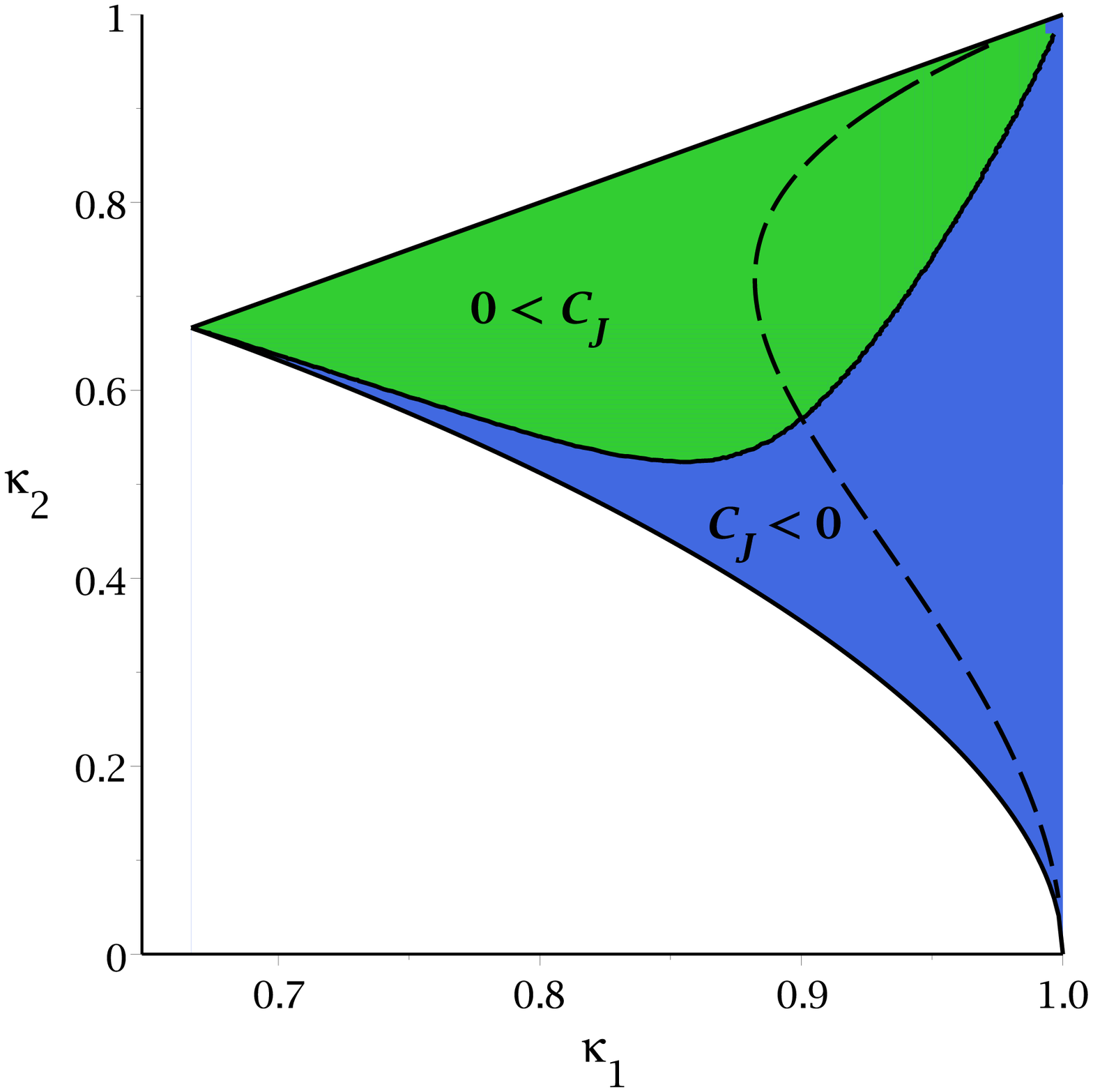}
 }
 \caption{Sign of the specific heat at constant angular momentum of the charged dilatonic black saturn in the fixed $Q$ ensemble. $C_J$ is positive in the green area and negative in the blue area. The balanced solutions can be found along the dashed curve.}
 \label{pic:saturn_cj}
\end{figure}

\begin{figure}[H]
 \centering
 \subfigure[$\beta=0$]{
  \includegraphics[width=5cm]{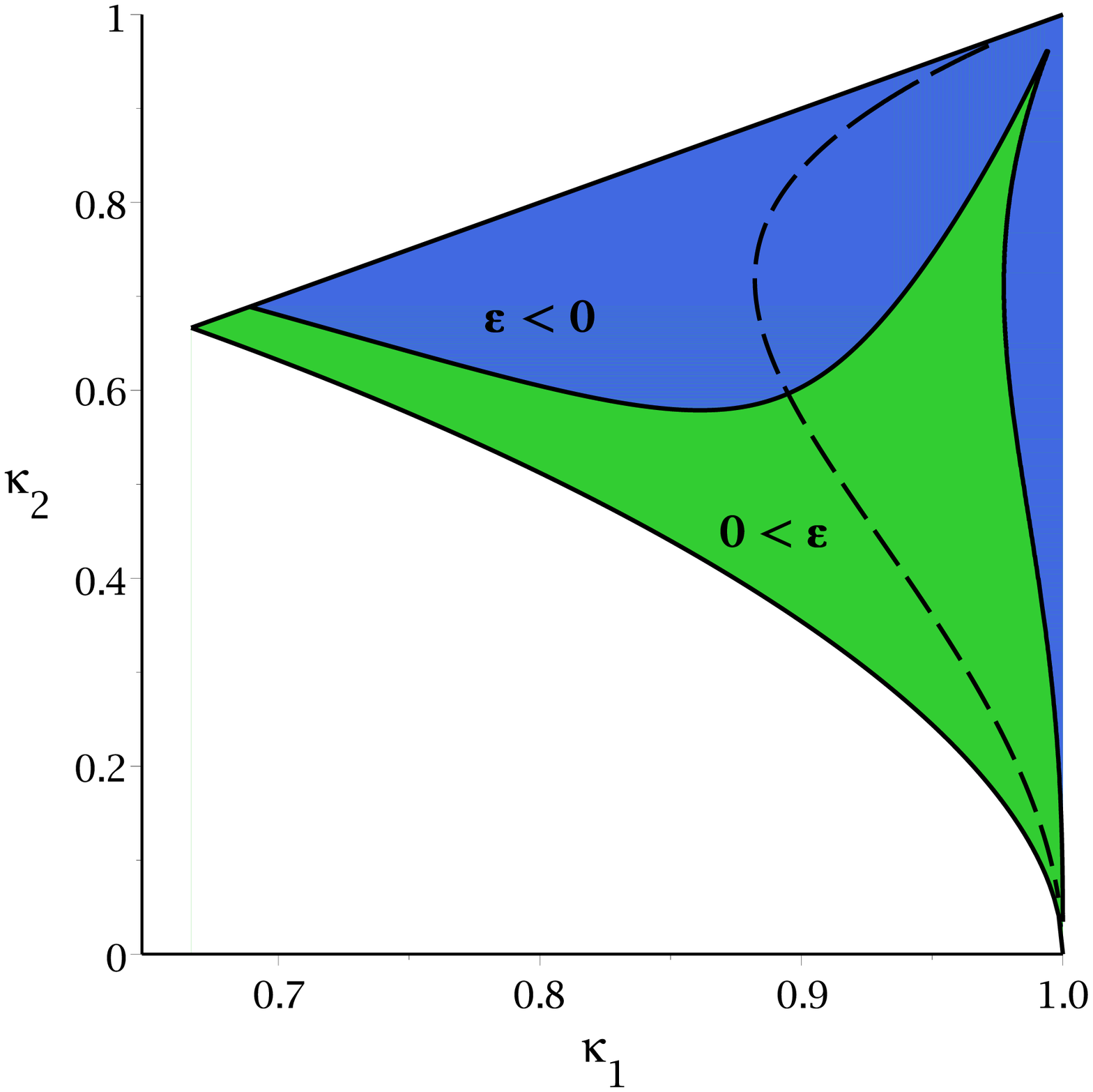}
 }
 \subfigure[$\beta=0.5$]{
  \includegraphics[width=5cm]{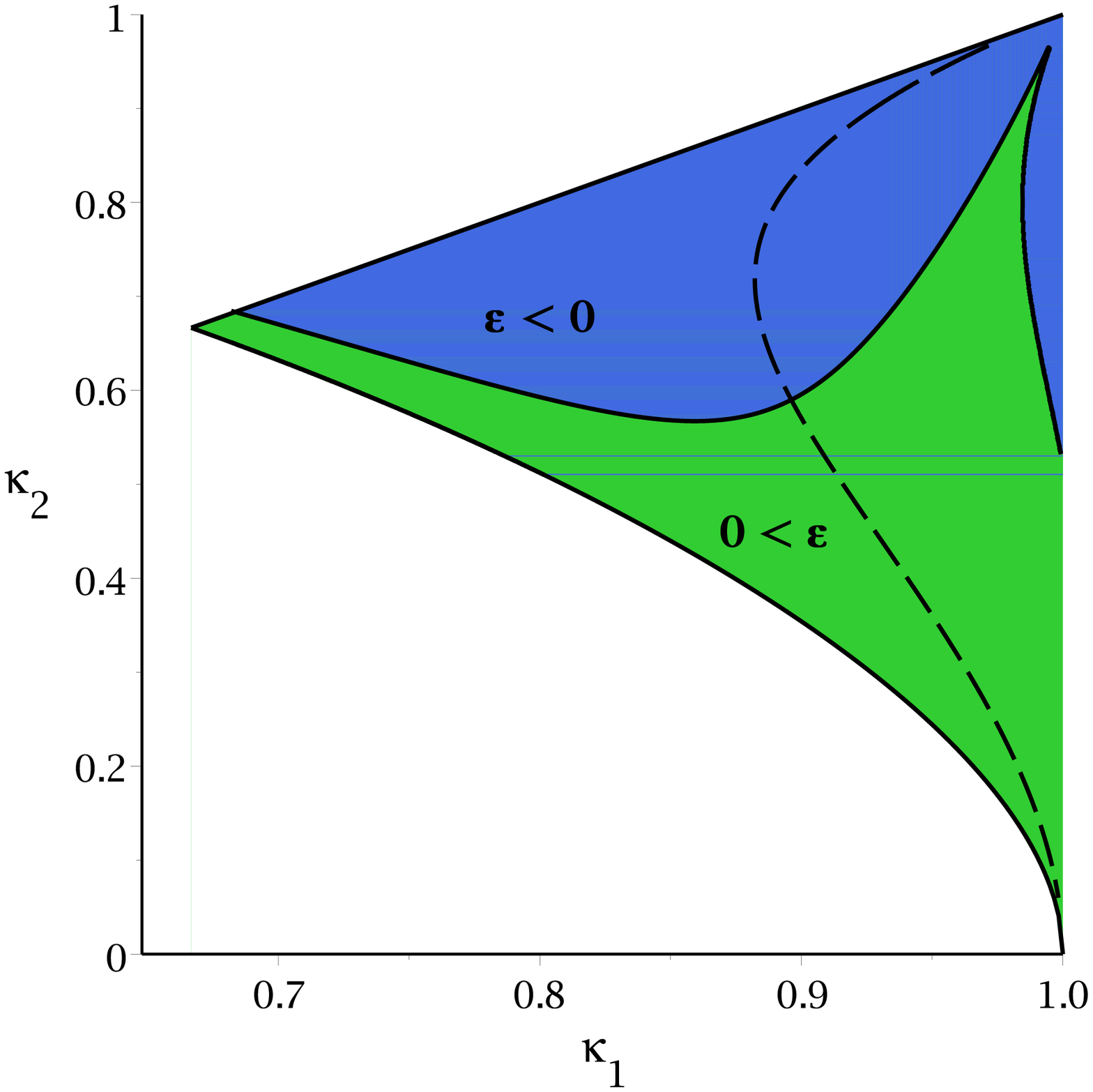}
 }
 \subfigure[$\beta=1$]{
  \includegraphics[width=5cm]{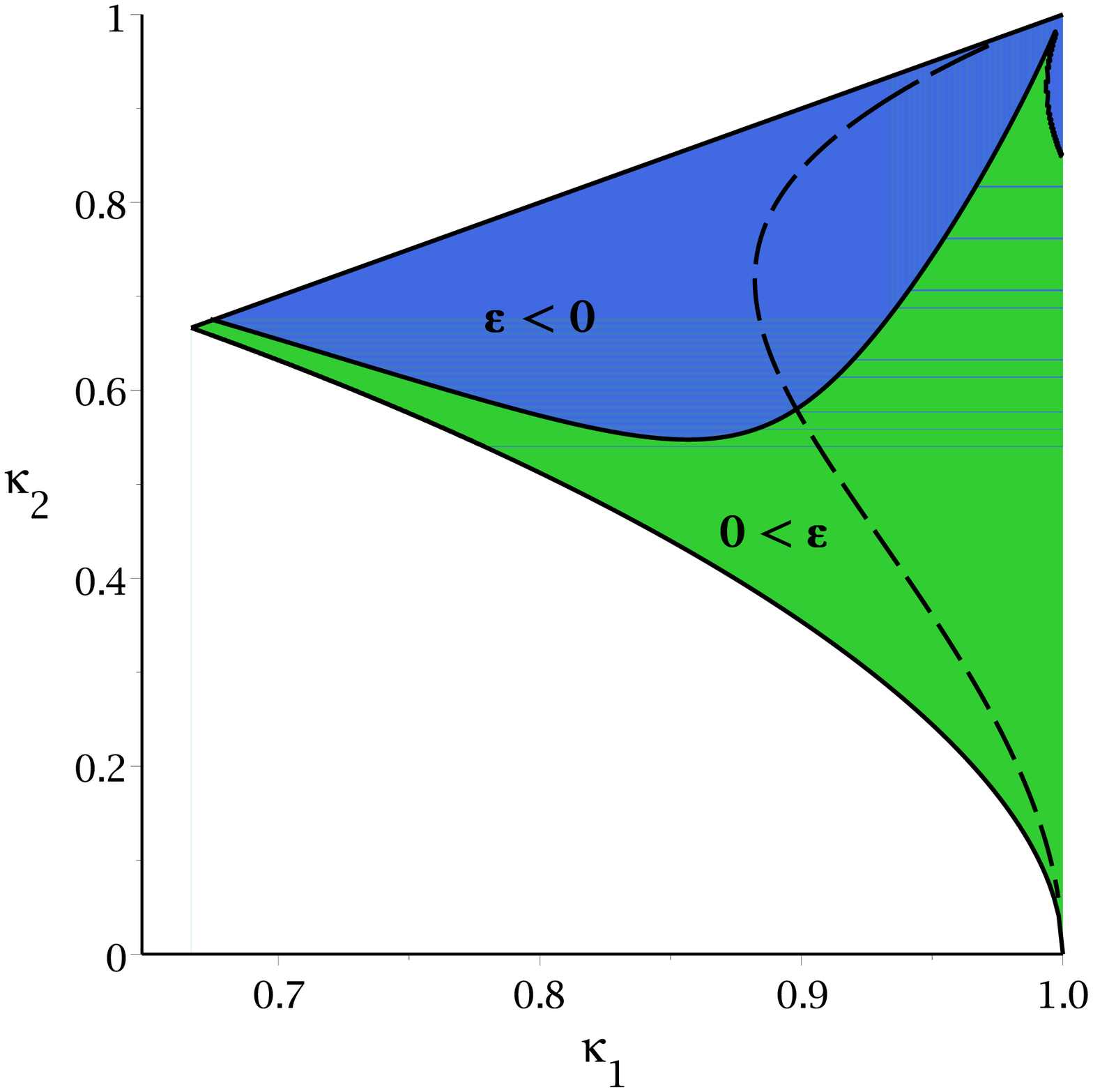}
 }
 \caption{Sign of the isothermal moment of inertia of the charged dilatonic black saturn in the fixed $Q$ ensemble. $\epsilon$ is positive in the green area and negative in the blue area. The balanced solutions can be found along the dashed curve.}
 \label{pic:saturn_eps}
\end{figure}

Figure \ref{pic:saturn_cj} and \ref{pic:saturn_eps} show $C_J$ and $\epsilon$ for different values of $\beta$ in the fixed $Q$ ensemble (the plots for $\beta=0$ also correspond to the fixed $\Psi_{\rm el}$ ensemble, since both ensembles are equal at $\beta=0$).

\clearpage

\section{Conclusion}

In this paper charged black rings and charged black saturns were constructed in five-dimensional Einstein-Maxwell-dilation theory with the Kaluza-Klein value of the dilaton coupling constant. The charged solutions have the same excess/deficit angle $\delta$ and the same mechanical moment of inertia as their neutral counterparts. The physical quantities of the charged solutions were calculated and related to the quantities of the corresponding neutral objects. It was shown, that the curve in the phase diagram of a charged dilatonic black ring or black saturn gets shifted to lower $a_H$ and $j^2$ as the charge parameter $\beta$ is increased.\\

Concerning the thermodynamical stability, the specific heat at constant angular momentum and the isothermal moment of inertia were considered. Both quantities have to be positive at the same time to ensure thermodynamical stability, but neither the charged dilatonic black ring nor the charged dilatonic black saturn are thermodynamically stable. Note that the neutral black ring and black saturn are also not thermodynamically stable \cite{Herdeiro:2010aq}.\\

For future work it would be interesting to construct charged dilatonic black rings and  black saturns with arbitrary values of the dilaton coupling contant, including the pure Einstein-Maxwell case, where the coupling constant vanishes. Subsequently, one would investigate their thermodynamic stability.\\

\begin{figure}[H]
 \centering
 \subfigure[Scaled electromagnetic term of the Smarr formula as a function of the scaled charge. The red dots correspond to the blackfold approximation of a charged black ring. The blue line corresponds to the analytic solution of a charged black ring/saturn.]{
  \includegraphics[width=7.5cm]{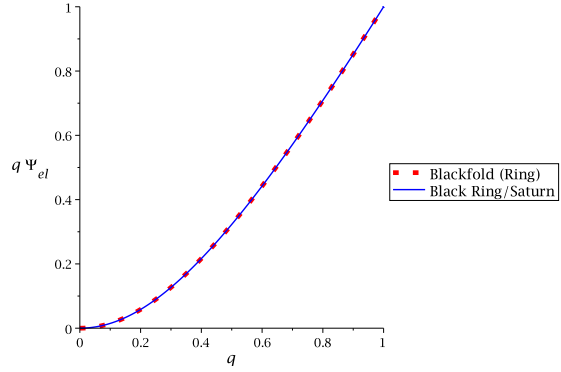}
 }

 \subfigure[Scaled horizon area term of the Smarr formula as a function of the scaled charge.The red dots correspond to the blackfold approximation of a charged black ring. The blue lines correspond to the analytic solution of a charged black ring. The ultraspinning limit is reached for $\nu\rightarrow 0$.]{
  \includegraphics[width=7.5cm]{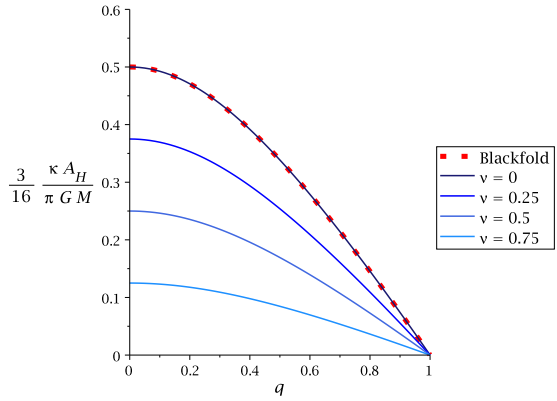}
 }
 \subfigure[Scaled angular momentum term of the Smarr formula as a function of the scaled charge.The red dots correspond to the blackfold approximation of a charged black ring. The blue lines correspond to the analytic solution of a charged black ring. The ultraspinning limit is reached for $\nu\rightarrow 0$.]{
  \includegraphics[width=7cm]{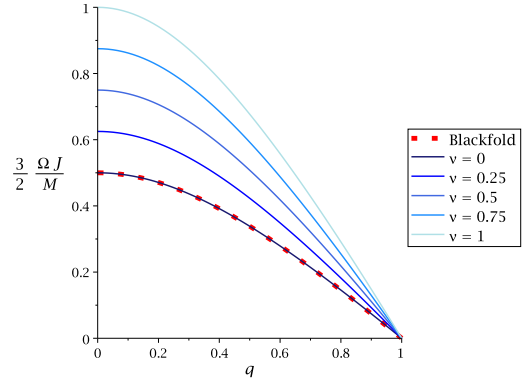}
 }

 \centering
 \subfigure[Scaled horizon area term of the Smarr formula as a function of the scaled charge. The red dots correspond to the blackfold approximation of a charged black ring. The blue lines correspond to the analytic solution of a charged black saturn. The ultraspinning limit is reached for $\kappa_2\rightarrow 0$.]{
  \includegraphics[width=8cm]{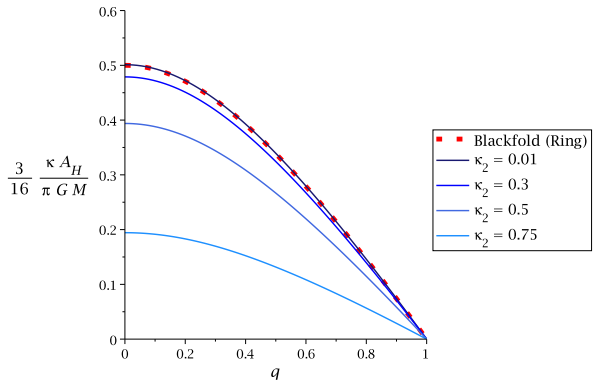}
 }
 \subfigure[Scaled angular momentum term of the Smarr formula as a function of the scaled charge.The red dots correspond to the blackfold approximation of a charged black ring. The blue lines correspond to the analytic solution of a charged black saturn. The ultraspinning limit is reached for $\kappa_2\rightarrow 0$.]{
  \includegraphics[width=7.5cm]{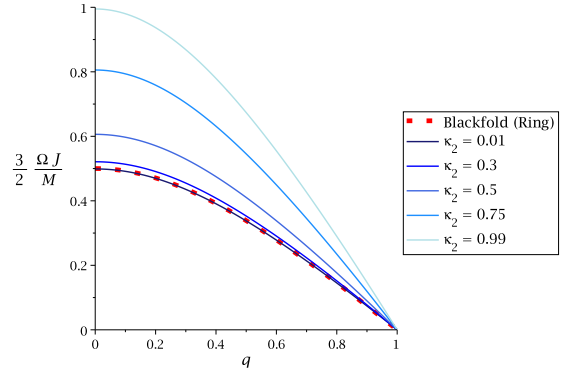}
 }
 \caption{Comparison of the blackfold approximation of the charged black ring with the analytic solution of the charged black ring and black saturn. In the ultraspinning limit the blackfold approximation of the black ring agrees well with the analytic black ring solution and also with the black saturn solution. }
\label{pic:blackfold}
\end{figure}

Since currently no analytical methods are available to obtain such more general charged solutions in closed form, one might resort to a combination of perturbative methods and numerical methods, to obtain a rather complete picture for such solutions. In the thin ring limit, where one has widely separate scales in the problem, the blackfold approach represents an excellent approximate method to study the properties of such solutions in five and more dimensions
\cite{Emparan:2007wm,Emparan:2009vd,Emparan:2009cs,Caldarelli:2010xz,Armas:2014bia}. In \cite{Caldarelli:2010xz} it was shown that the blackfold construction correctly reproduces all physical properties of the rotating charged dilatonic black ring. Interestingly the blackfold construction of a charged black ring also agrees very well with the charged black saturn in the ultraspinning limit. In a way this was expected since as $j$ is increased more mass is carried by the ring part of the black saturn and less by the central black hole, so that the $a_H$-$j$-curve of the black saturn asymptotes the black ring curve in the phase diagram. Figure \ref{pic:blackfold} shows a comparison of the blackfold approximation of the charged black ring with the analytic solution of the charged black ring and black saturn. The different (scaled) terms of the Smarr formula are plotted against the scaled charge $q=\frac{Q}{M}$.\\

In constrast, when the scales involved make the blackfold approach inappropriate, such as in the fat black ring regime, numerical methods can yield important new insights into the phase diagram of such higher dimensional black objects \cite{Kleihaus:2010hd,Kleihaus:2012xh,Dias:2014cia}.

\section{Acknowledgements}

I would like to thank Jutta Kunz, Niels Obers and Eugen Radu for helpful discussions, and also gratefully acknowledge support by the DFG, in particular, within the DFG Research Training Group 1620 ``Models of Gravity''.

%%%%%%%%%%%%%%%%%%%%%%%%%%%%%%%%%%%%%%%%%%

\bibliographystyle{unsrt}

\begin{thebibliography}{99}

%\cite{Emparan:2008eg}
\bibitem{Emparan:2008eg} 
  R.~Emparan and H.~S.~Reall,
  %``Black Holes in Higher Dimensions,''
  Living Rev.\ Rel.\  {\bf 11}, 6 (2008)
  [arXiv:0801.3471 [hep-th]].
  %%CITATION = ARXIV:0801.3471;%%

%\cite{Myers:1986un}
\bibitem{Myers:1986un}
  R.~C.~Myers and M.~J.~Perry,
  %``Black Holes In Higher Dimensional Space-Times,''
  Annals Phys.\  {\bf 172}, 304 (1986).
  %%CITATION = APNYA,172,304;%%

%\cite{Emparan:2001wn}
\bibitem{Emparan:2001wn} 
  R.~Emparan and H.~S.~Reall,
  %``A Rotating black ring solution in five-dimensions,''
  Phys.\ Rev.\ Lett.\  {\bf 88}, 101101 (2002)
  [hep-th/0110260].
  %%CITATION = HEP-TH/0110260;%%

%\cite{Pomeransky:2006bd}
\bibitem{Pomeransky:2006bd} 
  A.~A.~Pomeransky and R.~A.~Sen'kov,
  %``Black ring with two angular momenta,''
  hep-th/0612005.
  %%CITATION = HEP-TH/0612005;%%

%\cite{Elvang:2003yy}
\bibitem{Elvang:2003yy} 
  H.~Elvang,
  %``A Charged rotating black ring,''
  Phys.\ Rev.\ D {\bf 68}, 124016 (2003)
  [hep-th/0305247].
  %%CITATION = HEP-TH/0305247;%%

%\cite{Elvang:2004rt}
\bibitem{Elvang:2004rt} 
  H.~Elvang, R.~Emparan, D.~Mateos and H.~S.~Reall,
  %``A Supersymmetric black ring,''
  Phys.\ Rev.\ Lett.\  {\bf 93}, 211302 (2004)
  [hep-th/0407065].
  %%CITATION = HEP-TH/0407065;%%

%\cite{Elvang:2004ds}
\bibitem{Elvang:2004ds} 
  H.~Elvang, R.~Emparan, D.~Mateos and H.~S.~Reall,
  %``Supersymmetric black rings and three-charge supertubes,''
  Phys.\ Rev.\ D {\bf 71}, 024033 (2005)
  [hep-th/0408120].
  %%CITATION = HEP-TH/0408120;%%

%\cite{Elvang:2007rd}
\bibitem{Elvang:2007rd} 
  H.~Elvang and P.~Figueras,
  %``Black Saturn,''
  JHEP {\bf 0705}, 050 (2007)
  [hep-th/0701035].
  %%CITATION = HEP-TH/0701035;%%

%\cite{Iguchi:2007is}
\bibitem{Iguchi:2007is} 
  H.~Iguchi and T.~Mishima,
  %``Black di-ring and infinite nonuniqueness,''
  Phys.\ Rev.\ D {\bf 75}, 064018 (2007)
  [Erratum-ibid.\ D {\bf 78}, 069903 (2008)]
  [hep-th/0701043].

%\cite{Evslin:2007fv}
\bibitem{Evslin:2007fv} 
  J.~Evslin and C.~Krishnan,
  %``The Black Di-Ring: An Inverse Scattering Construction,''
  Class.\ Quant.\ Grav.\  {\bf 26}, 125018 (2009)
  [arXiv:0706.1231 [hep-th]].
  %%CITATION = ARXIV:0706.1231;%%

%\cite{Elvang:2007hs}
\bibitem{Elvang:2007hs} 
  H.~Elvang and M.~J.~Rodriguez,
  %``Bicycling Black Rings,''
  JHEP {\bf 0804}, 045 (2008)
  [arXiv:0712.2425 [hep-th]].

%\cite{Yazadjiev:2008ty}
\bibitem{Yazadjiev:2008ty} 
  S.~S.~Yazadjiev,
  %``Magnetized static black Saturn,''
  Phys.\ Rev.\ D {\bf 77}, 127501 (2008)
  [arXiv:0802.0784 [hep-th]].
  %%CITATION = ARXIV:0802.0784;%%

%\cite{Chng:2008sr}
\bibitem{Chng:2008sr} 
  B.~Chng, R.~B.~Mann, E.~Radu and C.~Stelea,
  %``Charging Black Saturn?,''
  JHEP {\bf 0812}, 009 (2008)
  [arXiv:0809.0154 [hep-th]].
  %%CITATION = ARXIV:0809.0154;%%

%\cite{Maison:1979kx}
\bibitem{Maison:1979kx} 
  D.~Maison,
  %``Ehlers-harrison Type Transformations For Jordan's Extended Theory Of Gravitation,''
  Gen.\ Rel.\ Grav.\  {\bf 10}, 717 (1979).
  %%CITATION = GRGVA,10,717;%%

%\cite{Maison:2000fj}
\bibitem{Maison:2000fj} 
  D.~Maison,
  %``Duality and hidden symmetries in gravitational theories,''
  Lect.\ Notes Phys.\  {\bf 540}, 273 (2000).
  %%CITATION = LNPHA,540,273;%%

%\cite{Chodos:1980df}
\bibitem{Chodos:1980df} 
  A.~Chodos and S.~L.~Detweiler,
  %``Spherically Symmetric Solutions in Five-dimensional General Relativity,''
  Gen.\ Rel.\ Grav.\  {\bf 14}, 879 (1982).
  %%CITATION = GRGVA,14,879;%%

%\cite{Dobiasch:1981vh}
\bibitem{Dobiasch:1981vh} 
  P.~Dobiasch and D.~Maison,
  %``Stationary, Spherically Symmetric Solutions of Jordan's Unified Theory of Gravity and Electromagnetism,''
  Gen.\ Rel.\ Grav.\  {\bf 14}, 231 (1982).
  %%CITATION = GRGVA,14,231;%%

%\cite{Gibbons:1985ac}
\bibitem{Gibbons:1985ac} 
  G.~W.~Gibbons and D.~L.~Wiltshire,
  %``Black Holes in Kaluza-Klein Theory,''
  Annals Phys.\  {\bf 167}, 201 (1986)
  [Erratum-ibid.\  {\bf 176}, 393 (1987)].
  %%CITATION = APNYA,167,201;%%

%\cite{Gibbons:1987ps}
\bibitem{Gibbons:1987ps} 
  G.~W.~Gibbons and K.~-i.~Maeda,
  %``Black Holes and Membranes in Higher Dimensional Theories with Dilaton Fields,''
  Nucl.\ Phys.\ B {\bf 298}, 741 (1988).
  %%CITATION = NUPHA,B298,741;%%

%\cite{Frolov:1987rj}
\bibitem{Frolov:1987rj} 
  V.~P.~Frolov, A.~I.~Zelnikov and U.~Bleyer,
  %``Charged Rotating Black Hole From Five-dimensional Point of View,''
  Annalen Phys.\  {\bf 44}, 371 (1987).
  %%CITATION = ANPYA,44,371;%%

%\cite{Breitenlohner:1987dg}
\bibitem{Breitenlohner:1987dg} 
  P.~Breitenlohner, D.~Maison and G.~W.~Gibbons,
  %``Four-Dimensional Black Holes from Kaluza-Klein Theories,''
  Commun.\ Math.\ Phys.\  {\bf 120}, 295 (1988).
  %%CITATION = CMPHA,120,295;%%

%\cite{Horne:1992zy}
\bibitem{Horne:1992zy} 
  J.~H.~Horne and G.~T.~Horowitz,
  %``Rotating dilaton black holes,''
  Phys.\ Rev.\ D {\bf 46}, 1340 (1992)
  [hep-th/9203083].
  %%CITATION = HEP-TH/9203083;%%

%\cite{Rasheed:1995zv}
\bibitem{Rasheed:1995zv} 
  D.~Rasheed,
  %``The Rotating dyonic black holes of Kaluza-Klein theory,''
  Nucl.\ Phys.\ B {\bf 454}, 379 (1995)
  [hep-th/9505038].
  %%CITATION = HEP-TH/9505038;%%

%\cite{Kunz:2006jd}
\bibitem{Kunz:2006jd} 
  J.~Kunz, D.~Maison, F.~Navarro-Lerida and J.~Viebahn,
  %``Rotating Einstein-Maxwell-dilaton black holes in D dimensions,''
  Phys.\ Lett.\ B {\bf 639}, 95 (2006)
  [hep-th/0606005].
  %%CITATION = HEP-TH/0606005;%%

%\cite{Kunduri:2004da}
\bibitem{Kunduri:2004da} 
  H.~K.~Kunduri and J.~Lucietti,
  %``Electrically charged dilatonic black rings,''
  Phys.\ Lett.\ B {\bf 609}, 143 (2005)
  [hep-th/0412153].
  %%CITATION = HEP-TH/0412153;%%

%\cite{Rocha:2013qya}
\bibitem{Rocha:2013qya} 
  J.~V.~Rocha, M.~J.~Rodriguez, O.~Varela and A.~Virmani,
  %``Charged black rings from inverse scattering,''
  Gen.\ Rel.\ Grav.\  {\bf 45}, 2099 (2013)
  [arXiv:1305.4969 [hep-th]].
  %%CITATION = ARXIV:1305.4969;%%

%\cite{Elvang:2003mj}
\bibitem{Elvang:2003mj} 
  H.~Elvang, R.~Emparan and ,
  %``Black rings, supertubes, and a stringy resolution of black hole nonuniqueness,''
  JHEP {\bf 0311}, 035 (2003)
  [hep-th/0310008].
  %%CITATION = HEP-TH/0310008;%%

%\cite{Rocha:2012vs}
\bibitem{Rocha:2012vs} 
  J.~V.~Rocha, M.~J.~Rodriguez and O.~Varela,
  %``An Electrically charged doubly spinning dipole black ring,''
  JHEP {\bf 1212}, 121 (2012)
  [arXiv:1205.0527 [hep-th]].
  %%CITATION = ARXIV:1205.0527;%%

%\cite{Feldman:2012vd}
\bibitem{Feldman:2012vd} 
  A.~Feldman and A.~A.~Pomeransky,
  %``Charged black rings in supergravity with a single non-zero gauge field,''
  JHEP {\bf 1207}, 141 (2012)
  [arXiv:1206.1026 [hep-th]].
  %%CITATION = ARXIV:1206.1026;%%

%\cite{Yazadjiev:2006ew}
\bibitem{Yazadjiev:2006ew} 
  S.~S.~Yazadjiev,
  %``Solution generating in 5D Einstein-Maxwell-dilaton gravity and derivation of dipole black ring solutions,''
  JHEP {\bf 0607}, 036 (2006)
  [hep-th/0604140].

%\cite{Yazadjiev:2007cd}
\bibitem{Yazadjiev:2007cd}
  S.~S.~Yazadjiev,
  %``Black Saturn with dipole ring,''
  Phys.\ Rev.\ D {\bf 76} (2007) 064011
  [arXiv:0705.1840 [hep-th]].

%\cite{Yazadjiev:2008pt}
\bibitem{Yazadjiev:2008pt} 
  S.~S.~Yazadjiev,
  %``5D Einstein-Maxwell solitons and concentric rotating dipole black rings,''
  Phys.\ Rev.\ D {\bf 78}, 064032 (2008)
  [arXiv:0805.1600 [hep-th]].

%\cite{Armas:2014rva}
\bibitem{Armas:2014rva} 
  J.~Armas and T.~Harmark,
  %``Constraints on the effective fluid theory of stationary branes,''
  arXiv:1406.7813 [hep-th].
  %%CITATION = ARXIV:1406.7813;%%

%\cite{Harmark:2004rm}
\bibitem{Harmark:2004rm} 
  T.~Harmark,
  %``Stationary and axisymmetric solutions of higher-dimensional general relativity,''
  Phys.\ Rev.\ D {\bf 70}, 124002 (2004)
  [hep-th/0408141].
  %%CITATION = HEP-TH/0408141;%%

%\cite{Liu:2010dq}
\bibitem{Liu:2010dq} 
  Z.~-X.~Liu and Z.~-Q.~Chen,
  %``Analysis of Unbalanced Black Ring Solutions within the Quasilocal Formalism,''
  Int.\ J.\ Mod.\ Phys.\ D {\bf 20}, 581 (2011)
  [arXiv:1010.4861 [hep-th]].
  %%CITATION = ARXIV:1010.4861;%%

%\cite{Herdeiro:2009vd}
\bibitem{Herdeiro:2009vd} 
  C.~Herdeiro, B.~Kleihaus, J.~Kunz and E.~Radu,
  %``On the Bekenstein-Hawking area law for black objects with conical singularities,''
  Phys.\ Rev.\ D {\bf 81}, 064013 (2010)
  [arXiv:0912.3386 [gr-qc]].
  %%CITATION = ARXIV:0912.3386;%%

%\cite{Herdeiro:2010aq}
\bibitem{Herdeiro:2010aq} 
  C.~Herdeiro, E.~Radu and C.~Rebelo,
  %``Thermodynamical description of stationary, asymptotically flat solutions with conical singularities,''
  Phys.\ Rev.\ D {\bf 81}, 104031 (2010)
  [arXiv:1004.3959 [gr-qc]].
  %%CITATION = ARXIV:1004.3959;%%

%\cite{Elvang:2007hg}
\bibitem{Elvang:2007hg} 
  H.~Elvang, R.~Emparan and P.~Figueras,
  %``Phases of five-dimensional black holes,''
  JHEP {\bf 0705}, 056 (2007)
  [hep-th/0702111].
  %%CITATION = HEP-TH/0702111;%%

%\cite{Emparan:2007wm}
\bibitem{Emparan:2007wm}
  R.~Emparan, T.~Harmark, V.~Niarchos, N.~A.~Obers and M.~J.~Rodriguez,
  %``The Phase Structure of Higher-Dimensional Black Rings and Black Holes,''
  JHEP {\bf 0710} (2007) 110
  [arXiv:0708.2181 [hep-th]].
  %%CITATION = JHEPA,0710,110;%%  

%\cite{Emparan:2009vd}
\bibitem{Emparan:2009vd}
  R.~Emparan, T.~Harmark, V.~Niarchos and N.~A.~Obers,
  %``New Horizons for Black Holes and Branes,''
  JHEP {\bf 1004}, 046 (2010)
  [arXiv:0912.2352 [hep-th]];
  %%CITATION = JHEPA,1004,046;%%

%\cite{Emparan:2009cs}
\bibitem{Emparan:2009cs}
  R.~Emparan, T.~Harmark, V.~Niarchos and N.~A.~Obers,
  %``Blackfolds,''
  Phys.\ Rev.\ Lett.\  {\bf 102} (2009) 191301
  [arXiv:0902.0427 [hep-th]].
  %%CITATION = PRLTA,102,191301; 

%\cite{Caldarelli:2010xz}
\bibitem{Caldarelli:2010xz} 
  M.~M.~Caldarelli, R.~Emparan and B.~Van Pol,
  %``Higher-dimensional Rotating Charged Black Holes,''
  JHEP {\bf 1104}, 013 (2011)
  [arXiv:1012.4517 [hep-th]].

%\cite{Armas:2014bia}
\bibitem{Armas:2014bia} 
  J.~Armas and T.~Harmark,
  %``Black Holes and Biophysical (Mem)-branes,''
  arXiv:1402.6330 [hep-th].

%\cite{Kleihaus:2010hd}
\bibitem{Kleihaus:2010hd} 
  B.~Kleihaus, J.~Kunz and K.~Schn\"ulle,
  %``Charged Balanced Black Rings in Five Dimensions,''
  Phys.\ Lett.\ B {\bf 699}, 192 (2011)
  [arXiv:1012.5044 [hep-th]].

%\cite{Kleihaus:2012xh}
\bibitem{Kleihaus:2012xh} 
  B.~Kleihaus, J.~Kunz and E.~Radu,
  %``Black rings in six dimensions,''
  Phys.\ Lett.\ B {\bf 718}, 1073 (2013)
  [arXiv:1205.5437 [hep-th]].

%\cite{Dias:2014cia}
\bibitem{Dias:2014cia} 
  O.~J.~C.~Dias, J.~E.~Santos and B.~Way,
  %``Rings, Ripples, and Rotation: Connecting Black Holes to Black Rings,''
  arXiv:1402.6345 [hep-th].


\end{thebibliography}

\end{document}